\documentclass[10pt,prl,twocolumn,letterpaper,showpacs,groupedaddress,superscriptaddress,nofootinbib,floatfix,preprintnumbers,tightenlines]{revtex4}
\usepackage{amssymb,amsmath,graphicx,xcolor,cancel,dcolumn,tabularx}
\usepackage{bm,comment}
\usepackage{color,float}

\newcolumntype{d}[1]{D{.}{.}{1.4}}

\newcommand{\lsim}{\raisebox{-0.7ex}{$\stackrel{\textstyle <}{\sim}$ }}

\begin{document}

\begin{figure}
  \vskip -1.5cm
  \leftline{\includegraphics[width=0.18\textwidth]{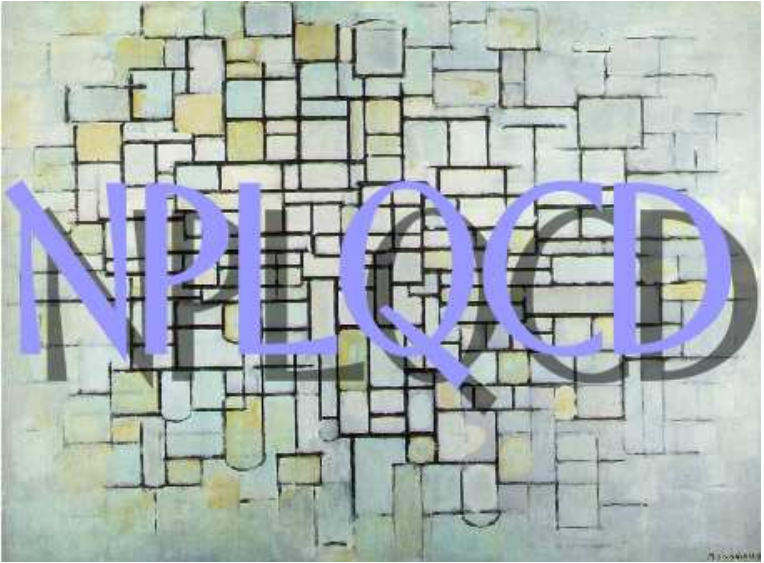}}
\end{figure}

\vspace{-1cm}
\title{Scalar, axial and tensor interactions of light nuclei from lattice QCD }

\author{Emmanuel Chang}
\noaffiliation{}

\author{Zohreh Davoudi}
\affiliation{Department of Physics, University of Maryland, College Park, MD 20742, USA}
\affiliation{Kavli Institute for Theoretical Physics, University of California, Santa Barbara, CA 93106, USA}

\author{William Detmold}
\affiliation{Center for Theoretical Physics, Massachusetts Institute of Technology, Cambridge, MA 02139, USA}
\affiliation{Kavli Institute for Theoretical Physics, University of California, Santa Barbara, CA 93106, USA}

\author{Arjun S. Gambhir}
\affiliation{Nuclear and Chemical Sciences Division,
	Lawrence Livermore National Laboratory, Livermore, CA, 94550}
\affiliation{Nuclear Science Division, Lawrence Berkeley National Laboratory, Berkeley, CA, 94720}

\author{Kostas Orginos}
\affiliation{Department of Physics, College of William and Mary, Williamsburg, VA 23187-8795, USA}
\affiliation{Jefferson Laboratory, 12000 Jefferson Avenue, 
	Newport News, VA 23606, USA}

\author{Martin J. Savage}
\affiliation{Institute for Nuclear Theory, University of Washington, Seattle, WA 98195-1550, USA}
\affiliation{Kavli Institute for Theoretical Physics, University of California, Santa Barbara, CA 93106, USA}

\author{Phiala E. Shanahan}
\affiliation{Department of Physics, College of William and Mary, Williamsburg, VA 23187-8795, USA}
\affiliation{Jefferson Laboratory, 12000 Jefferson Avenue, Newport News, VA 23606, USA}
\affiliation{Kavli Institute for Theoretical Physics, University of California, Santa Barbara, CA 93106, USA}

\author{Michael L. Wagman}
\affiliation{Center for Theoretical Physics, Massachusetts Institute of Technology, Cambridge, MA 02139, USA}
\affiliation{Kavli Institute for Theoretical Physics, University of California, Santa Barbara, CA 93106, USA}

\author{Frank Winter}
\affiliation{Jefferson Laboratory, 12000 Jefferson Avenue, 
	Newport News, VA 23606, USA}

\collaboration{NPLQCD Collaboration}

\date{\today}

\preprint{INT-PUB-17-054}
\preprint{MIT-CTP/4967}
\preprint{JLAB-THY-17-2607}
\preprint{UMD-PP-017-35}


\begin{abstract}
Complete flavor decompositions of the  matrix elements of the scalar, axial and tensor 
currents in the proton, deuteron, diproton and $^3$He 
at SU(3)-symmetric values of the quark masses corresponding to a pion mass $m_\pi\sim806$~MeV are determined using lattice quantum chromodynamics. 
At the physical quark masses, the scalar interactions constrain mean-field models of nuclei and the low-energy interactions of nuclei with potential dark matter candidates. 
The axial and tensor interactions of nuclei constrain their spin content,  integrated transversity 
and the quark contributions to their electric dipole moments. 
External fields are used to directly access the quark-line connected matrix elements of quark bilinear operators, 
and a  combination of stochastic estimation techniques is used to determine the disconnected sea-quark contributions. The calculated matrix elements differ from, and are typically smaller than, naive single-nucleon estimates.
 Given the particularly large, $\mathcal{O}(10\%)$, size of nuclear effects in the scalar matrix elements, contributions from correlated multi-nucleon effects should be quantified in the analysis of dark matter direct-detection experiments using nuclear targets.
\end{abstract}

\maketitle

Understanding the spin and flavor structure of nuclei at the level of quarks and gluons is essential to the interpretation of many 
searches for  beyond the Standard Model (BSM) physics. 
The simplest aspects of the structure of nuclei are revealed through their static responses to 
external probes.
Vector charges of a nucleus are constrained by symmetries and define the number of valence quarks of a given flavor, while
 matrix elements (MEs) of the axial currents encode the spin carried by quarks and 
gluons ~\cite{Ashman:1987hv,Ji:1996ek,Jaffe:1989jz} 
and play a central role in 
weak-interaction processes including  single and double-$\beta$ decay. 
While difficult to probe experimentally, the
(renormalization-scale--dependent) scalar and tensor MEs provide important theoretical input for the interpretation of  results from dark matter 
direct detection experiments~\cite{Undagoitia:2015gya} 
and searches for new physics in precision spectroscopy~\cite{Delaunay:2016brc,Delaunay:2017dku}.
{\color{black} Tensor MEs determine the quark electric dipole moment (EDM) contributions to nuclear EDMs through the dimension-five CP-odd operator $\overline{q}\sigma_{\mu\nu}q\ \widetilde F^{\mu\nu}$ (where $\widetilde F^{\mu\nu}=\frac{1}{2}\epsilon^{\mu\nu\rho\sigma}F_{\rho\sigma}$ is the dual of the electromagnetic field strength tensor $F^{\mu\nu}$ and $q$ is the quark field)} and are necessary to interpret corresponding searches for BSM CP violation~\cite{Engel:2013lsa,Yamanaka:2016umw,Yamanaka:2017mef,Chupp:2017rkp}.


In interpreting intensity-frontier searches for new physics using nuclear targets, it is important to consider multi-nucleon effects in nuclear MEs. 
For axial MEs, relevant for Gamow-Teller (GT) transitions, experimental measurements generally differ substantially
from naive single-nucleon (NSN) estimates using
nuclear ground states with non-interacting nucleons
occupying only the lowest shell-model states~\cite{Buck:1975ae,Krofcheck:1985fg,Chou:1993zz,Brown:1978zz,Wildenthal:1983zz}.
Phenomenologically, nuclear shell-model calculations of $\beta$-decay rates using quenched values of the nucleon axial coupling are known to agree better with experimental values~\cite{Brown:1978zz,Wildenthal:1983zz,MartinezPinedo:1996vz,Kumar:2016snu,Deppisch:2016rox}.
For light nuclei with $A \le$ 10, recent Green function Monte Carlo calculations of GT MEs using chiral currents and potentials~\cite{Schiavilla:1998je,Krebs:2016rqz,Baroni:2016xll,De-Leon:2016wyu,Pastore:2017uwc} have shown that experimental values of axial MEs can be reproduced by including correlated two-nucleon effects, constrained by experimental  observations in few-body  systems. 
In larger nuclei, multi-body nuclear effects make such calculations significantly more challenging,  (see, e.g., Refs.~\cite{Shafer:2016etk,Perez:2017ksf} for recent progress).
For the scalar and tensor currents, chiral effective field theories (EFTs) have also been used to organize multi-nucleon effects in nuclear MEs~\cite{Engel:1992bf,Prezeau:2003sv,Ellis2008,Ellis2009,Giedt2009,Hill2011,Menendez:2011qq,Ellis2012,Underwood2012,Korber:2017ery,Cirigliano:2012pq,Fitzpatrick:2012ix,Klos:2013rwa,Hoferichter:2015ipa,Hoferichter:2016nvd,Bishara:2016hek,Hoferichter:2017olk}. 
In contrast to the axial case, these effects are not well constrained by experiment, but could be determined by matching to ab-initio Standard Model calculations of few-body matrix elements using lattice quantum chromodynamics (lattice QCD). 
In principle, lattice QCD can be used to calculate the nuclear MEs of interest directly from quark and gluon interactions. While such calculations are extremely challenging, they have now been performed for $A\le 4$, albeit without fully controlled uncertainties, and will be improved and extended to larger nuclei through advances in algorithms and growth of computational resources.

In this Letter, a first-principles lattice QCD study of forward matrix elements of the scalar, axial and tensor currents,
henceforth referred to as  ``charges'',
in the nucleon and light nuclei up to 
atomic number $A=3$ is presented, at unphysical values of the quark masses.
These nuclear matrix elements are seen to deviate from the NSN estimates, with particularly large deviations for the scalar current. 
As  many  theories  of  dark matter couple to the SM through scalar exchange, it is important to quantify these potentially large effects in the interpretation of dark matter direct detection experiments.

\vspace*{3mm}
{\it Lattice QCD methodology}: 
The numerical calculations presented here are performed using one ensemble of gauge-field configurations generated with a clover-improved 
fermion action~\cite{Sheikholeslami:1985ij} and a L\"uscher-Weisz gauge action~\cite{Luscher:1984xn} with $N_f=3$ degenerate light-quark flavors. 
The quark masses are tuned to produce a pion of mass $m_\pi\sim 806~{\rm MeV}$. 
The spacetime volume of the ensemble is $L^3\times T=32^3\times48$, and the lattice spacing is $a\sim 0.145~{\rm fm}$. Further details of the ensemble are given in Ref.~\cite{Beane:2012vq}. 

The matrix elements of the scalar, axial and tensor currents are determined from lattice QCD correlation functions calculated 
on each gauge-field configuration. 
These correlation functions separate into two pieces; one in which the quarks interacting with the operator are connected to the hadronic source and sink, 
and one in which they are not, referred to as quark-line connected and disconnected contributions, respectively. 
The quark-line connected contributions to the matrix elements are determined using the fixed-order background-field approach introduced in 
Ref.~\cite{Savage:2016kon} and discussed in detail for the axial case in Refs.~\cite{Bouchard:2016heu,Tiburzi:2017iux,Shanahan:2017bgi}. 
Here, Dirac bilinears 
$\overline{q} q$, $\overline{q}\gamma_3 \gamma_5 q$, 
and $\overline{q}\gamma_1 \gamma_2q$ for $q=u,d$
are used to couple to the scalar, axial and tensor fields, respectively.
Correlation functions are constructed with seven different values of the field strengths for proton ($p$), deuteron ($d)$, diproton  ($pp$), and ${}^3$He states. 
These correlation functions are built from quark propagators originating from a smeared source and having either a smeared (SS) or point (SP) sink \cite{Tiburzi:2017iux}. Matrix elements are extracted from the linear responses of the correlation functions to the external fields as detailed in Refs.~\cite{Savage:2016kon,Tiburzi:2017iux,Shanahan:2017bgi}, using correlated one- and two-state fits. 
As an illustration, 
fits determining the connected isoscalar scalar matrix element in  $^3$He are shown in Fig.~\ref{fig:connected}.
Fits for all states, as well as details of the statistical sampling, are provided in the supplementary material.
It is important to note that at the SU(3)-symmetric value of the quark masses used in this study, 
all of the states considered are bound ground states {\color{black} with binding energies significantly larger than those in nature}~\cite{Orginos:2015aya,Beane:2012vq,Wagman:2017tmp,Beane:2017edf}, and are spatially compact 
with respect to the lattice volume. Finite-volume effects in the matrix elements are therefore exponentially small in $\gamma L$, where $\gamma$ is the binding momentum of the system. 
\begin{figure}
	\centering
	\includegraphics[width=0.9\columnwidth]{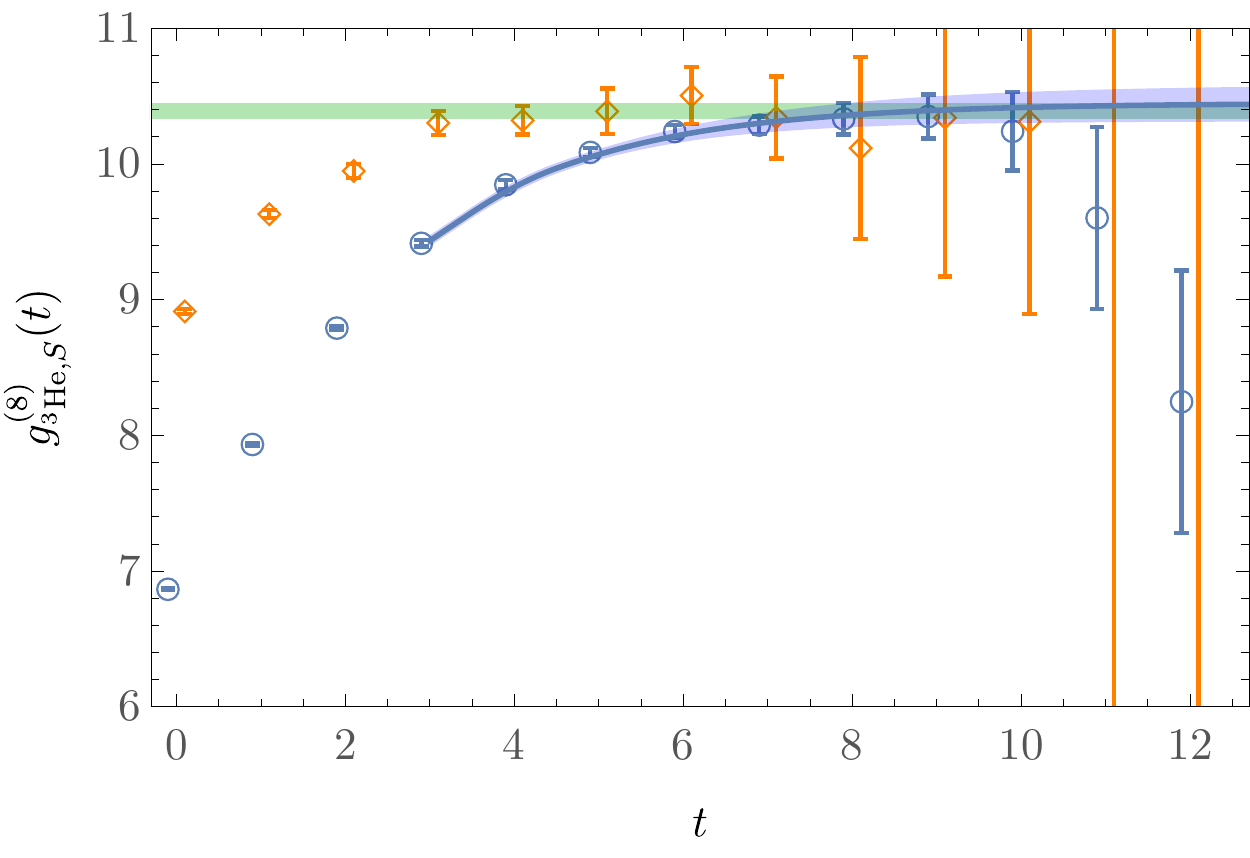}
	\caption{\label{fig:connected} 
	The bare effective matrix element of the connected isoscalar $^3$He scalar charge, {\color{black} $g^{(8)}_{{}^3{\rm He},S}=\langle {}^3{\rm He} |\overline{q} \Lambda^{(8)} q | {}^3{\rm He}\rangle$}, 
	where the blue circles and orange diamonds denote SP and SS results, respectively. 
	The blue band illustrates a correlated two-state
	fit of the form {\color{black} $\sim A + B\ e^{-\Delta\, t}$,} to the SP correlation functions, while the green band denotes the final fit result with 
	combined statistical and systematic uncertainties. 
	}
\end{figure}

\begin{figure}
	\centering 
	\includegraphics[width=0.9\columnwidth]{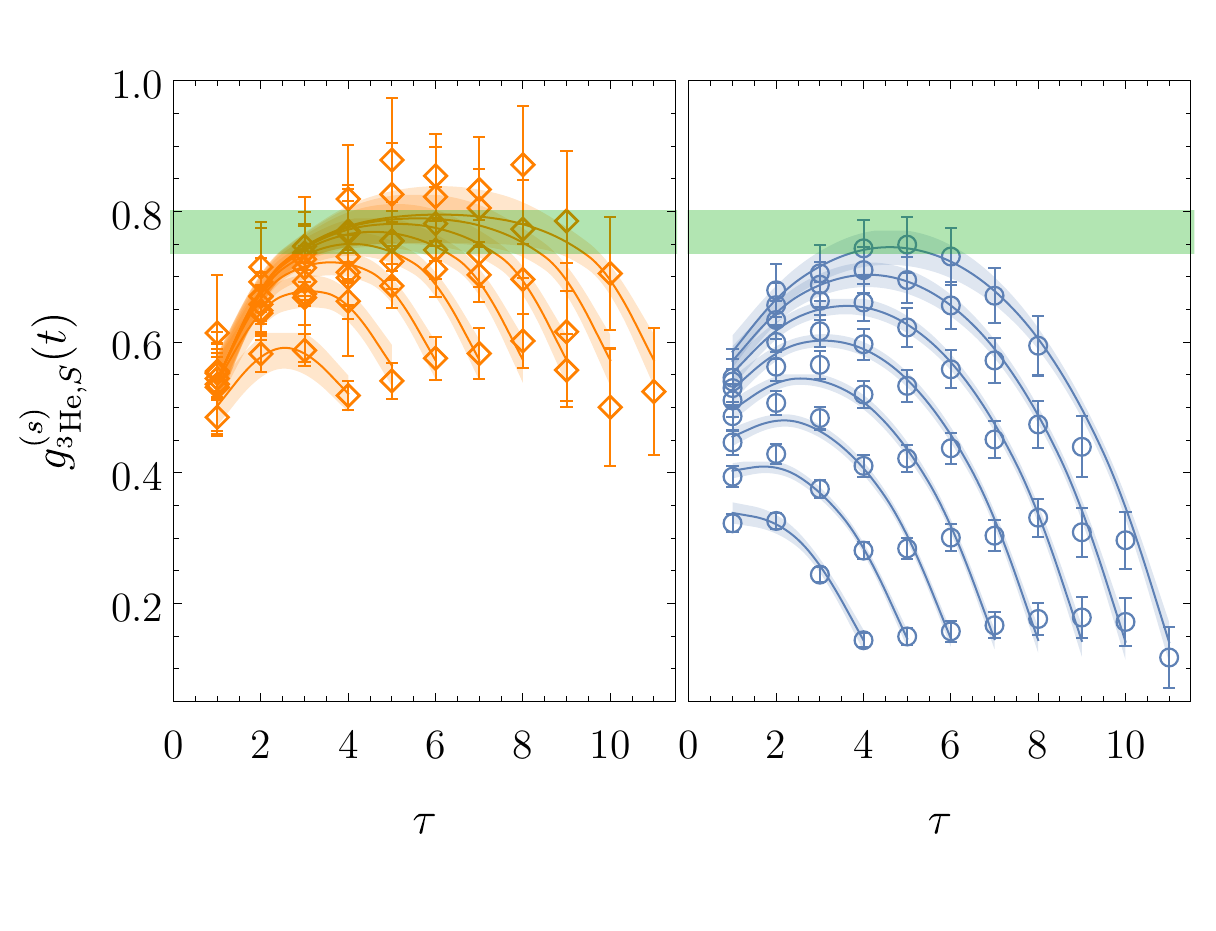}
	\caption{\label{fig:disco} 
	The strange-quark (disconnected) scalar matrix element in $^3$He,
	{\color{black} $g^{(s)}_{{}^3{\rm He},S}=\langle {}^3{\rm He} |\overline{s} s | {}^3{\rm He}\rangle$}. The left (right) panel shows results obtained using the SS (SP) correlation functions for a range of current insertion times $\tau$ and sink insertion times. 
	The  green band corresponds to the extracted matrix element determined as described in the text, 
	and the blue (orange) curves and bands illustrate a correlated two-state fit to the data shown as discussed in the supplementary material.
	}
\end{figure}

Calculation of the quark-line disconnected contributions to each matrix element requires all-to-all propagators 
and it is not feasible to compute these objects exactly for the lattice volume used in this work. 
Instead, the requisite traces are estimated stochastically~\cite{Hutchinson_90, Dong:1993pk, Neff:2001zr, Foley:2005ac, Babich:2007jg} using hierarchical probing~\cite{Stathopoulos:2013aci} and 
singular-value deflation~\cite{Gambhir:2016uwp, Gambhir:2016jul} (see Ref.~\cite{Gambhir:2016uwp,ArjunThesis} for complete details). 
{\color{black} The uncertainties from the stochastic sampling are included in the statistical uncertainties that are presented.}
The disconnected contributions are subsequently correlated with SS and SP two-point correlation functions with the relevant quantum numbers to 
construct the three-point correlation functions. After subtraction of 
the contributions in which the current insertions and the 
two-point functions are uncorrelated, the ratios of the three-point functions to the two-point functions 
are formed. Finally, the ground-state matrix elements are extracted using correlated two-state fits to the sink and operator-insertion time-dependence. 
Bare results for the $^3$He disconnected scalar matrix element and the corresponding fits are shown in Fig.~\ref{fig:disco}. 
Further technical details of the statistical sampling of the correlation functions, as well as figures showing the analysis of the other disconnected matrix elements, 
are presented in the supplementary material.

Combining the quark-line connected and disconnected contributions to the matrix elements allows  a complete flavor decomposition of the scalar, axial,
and tensor charges in the light nuclei studied. 
The bare lattice operators are renormalized using the flavor-nonsinglet renormalization constants $Z_S=0.823(16)$, $Z_A=0.879(12)$,  and $Z_T=0.889(16)$, 
determined in Ref.~\cite{Yoon:2016jzj} from ensembles with the same action. 
For the scalar and tensor operators, results are presented in the $\overline{\rm MS}$ scheme at a renormalization scale of $\mu=2$ GeV. 
The isovector charges are free from significant operator mixing, while the isoscalar matrix elements are only determined up to mixings with gluon operators, 
which are not computed.  
For the isoscalar scalar and tensor operators, this is a small effect~\cite{Alexandrou:2016ekb,Alexandrou:2017qyt}, while 
for the isoscalar axial charges, mixing through {\color{black} the flavor-singlet chiral anomaly} is potentially significant. 
Nevertheless, to leading order in the strong coupling, these renormalization factors, mixing, and the renormalization-scale--dependence 
of the scalar and tensor charges,  cancel in ratios of nuclear matrix elements to the matrix elements of the same operators in the 
proton~\cite{Winter:2017bfs}. 
These ratios, and their differences from the expectations for non-interacting collections of nucleons, 
encode nuclear effects and are the primary focus of this work.

\vspace*{3mm}
{\it Proton and nuclear charges}: 
The renormalized scalar, axial and tensor charges of the proton,  deuteron,  diproton and $^3$He  are given in Table~\ref{tab:ren} 
(the bare charges are presented in the supplementary material). 
Results are given in the basis of flavor matrices
\footnote{For notational convenience, a non standard normalization of the  flavor matrices is used. These are related to the Gell-Mann matrices $\lambda_i$ as  $\Lambda^{(3)}=\lambda_3$ and $\Lambda^{(8)}=\sqrt{3}\lambda_8$.} $\Lambda^{(3)}\equiv{\rm diag}(1,-1,0)$, $\Lambda^{(8)}\equiv{\rm diag}(1,1,-2)$, 
and the identity $\Lambda^{(0)}\equiv{\rm diag}(1,1,1)$, with the charges  labeled as $g_X^{(3,8,0)}$   
respectively, where $X=S,A,T$ indicates the Dirac structure.  
Since the calculations are performed in the limit of SU(3)$_f$ flavor symmetry, the disconnected contributions cancel in both $g_X^{(3)}$ and $g_X^{(8)}$. 
{\color{black} The disconnected contributions (equivalently for the non-strange hadrons considered here, the strangeness contributions) 
are defined by the difference $g_X^{(\rm disc.)} = g_X^{(s)} = \left(g_X^{(0)} -g_X^{(8)}\right)/3$. }
For convenience, these contributions are given separately in Table~\ref{tab:ren}. 
\begin{table}
	\begin{ruledtabular}
		\begin{tabular}{c|cccc}
			& $p$ & $d$ & $pp$ & ${}^3$He\\ 
			\hline
			$g_S^{(0)}$	&3.65(7) &7.20(15)& 7.22(15) & 10.4(2) \\ 
			$g_S^{(3)}$	&0.78(2) & - & 1.55(4) & 0.77(2) \\ 
			$g_S^{(8)}$	& 2.94(6)& 5.84(12)& 5.86(12) & 8.55(18)  \\ 
			$g_S^{(s)}$	& 0.234(8)& 0.45(2) & 0.45(2) & 0.63(3) 
			\\	\hline
			$g_A^{(0)}$	& 0.634(9) &1.26(2) & - & 0.63(1)  \\ 
			$g_A^{(3)}$	& 1.14(2) & - & - & 1.13(2) \\ 
			$g_A^{(8)}$	& 0.633(9) & 1.25(2) & - & 0.625(9)  \\ 
			$g_A^{(s)}$	& 0.0002(6) & 0.001(1) & - & 0.003(2)   \\ \hline
			$g_T^{(0)}$	& 0.684(12) & 1.36(2) & - & 0.678(12)  \\ 
			$g_T^{(3)}$	& 1.12(2) & - & - & 1.12(3) \\ 
			$g_T^{(8)}$	& 0.684(12) & 1.36(2) & - & 0.676(12)  \\ 
			$g_T^{(s)}$	& 0.00007(13) & 0.0002(2) & - & 0.0004(4)  \\ 
		\end{tabular}
	\end{ruledtabular} 
	\caption{
	The renormalized scalar, axial and tensor charges of the proton and light nuclei at a renormalization scale 
	of $\mu=2$ GeV in the $\overline{\rm MS}$ scheme, neglecting mixing with gluonic operators. 
{\color{black} 	Specifically, for a nucleus $A$, $g_S^{(i)}=\langle A | \overline{q} \Lambda^{(i)}q|A\rangle$,  $g_A^{(i)}=\langle A | \overline{q} \gamma_3\gamma_5 \Lambda^{(i)}q|A\rangle$,  $g_T^{(i)}=\langle A | \overline{q} \sigma_{12}\Lambda^{(i)}q|A\rangle$, with the flavor structures $\Lambda^{(i)}$ defined in the text. }
	Statistical uncertainties, the systematic uncertainties arising from choices of fit procedure, and the uncertainties of the renormalization constants, 
	have been combined in quadrature.  
	}
	\label{tab:ren}
\end{table}

\begin{figure}
	\centering
	\includegraphics[width=0.9\columnwidth]{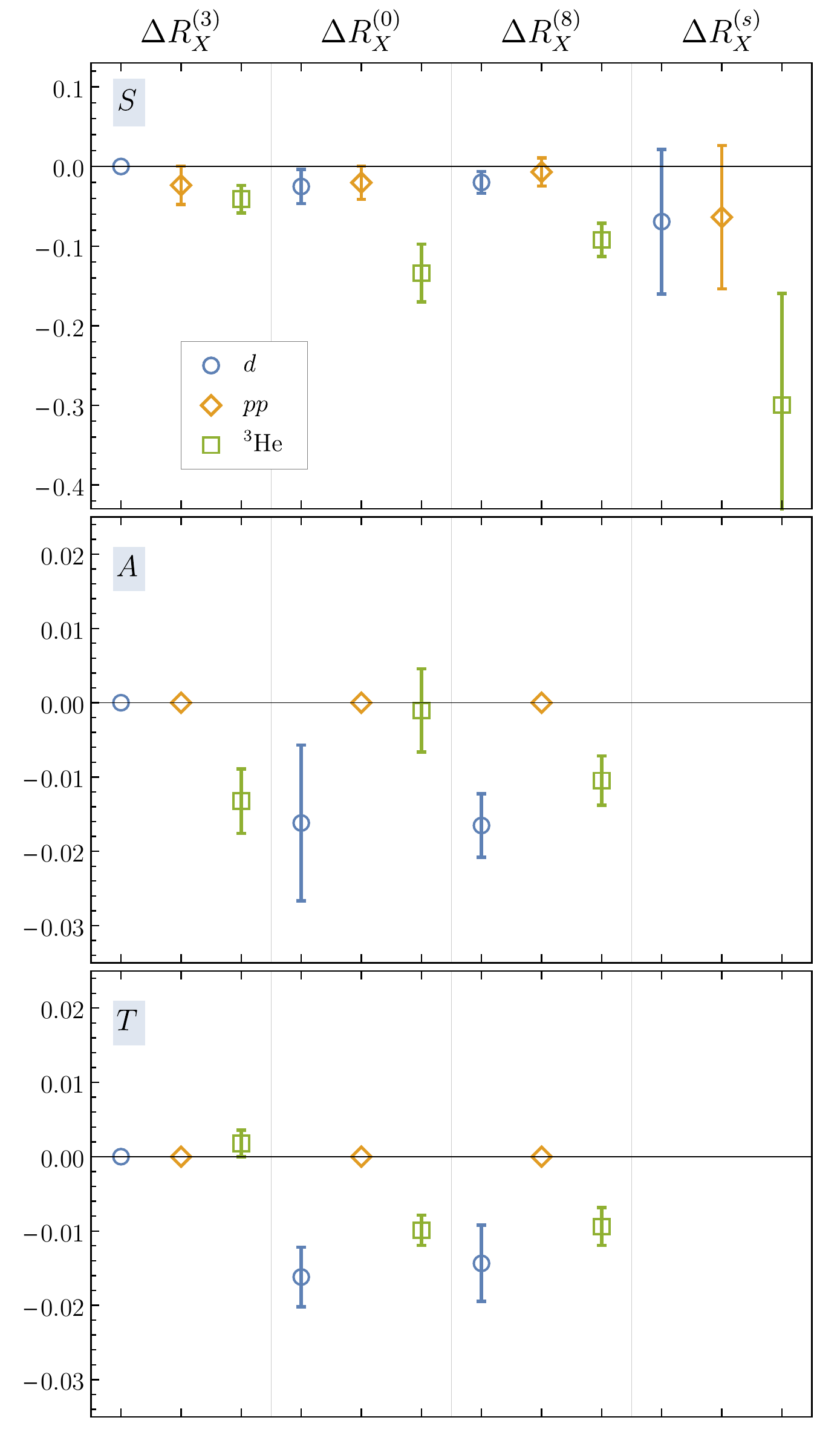}
	\caption{\label{ratsummary} 
	The calculated values of $\Delta R_X^{(f)}$ for the deuteron (circles), diproton (diamonds) and $^{3}{\rm He}$ (squares) 
	to those in the proton. 
	The  panels display the results obtained for the  scalar (top), axial (middle) and tensor (bottom)  interactions, and the 
	 columns within the panels display results for the different flavor structures of the currents, 
	 as indicated at the top of the figure. 
	In each case, the  statistical and systematic uncertainties have been combined in quadrature. 
	The points exactly at zero are constrained to vanish by spin and/or isospin symmetry, 
	while ratios are not given for the strange quark axial and tensor charges as both the numerators and denominator 
	are consistent with zero.
	}
\end{figure}

The ratios of the  charges in a nucleus $A$ to those in the proton, $R_X^{(f)}(A)=g_X^{(f)}(A)/g_X^{(f)}(p)$, can be compared
with the NSN estimates, defined previously, which are determined entirely by the baryon number, 
isospin, and spin quantum numbers.
Most sources of systematic uncertainty in these calculations, 
such as lattice spacing and finite volume effects,  cancel to a significant extent in these ratios~\cite{Winter:2017bfs}.
Figure~\ref{ratsummary} summarizes the differences $\Delta R_X^{(f)}(A)=R_X^{(f)}(A)- R_X^{(f)}(A)_{\rm NSN}$, which highlight the effects of nuclear interactions and correlations
on the charges, and 
present a coherent picture of medium effects in light nuclei at $m_\pi \sim 806$ MeV---the central results of this Letter. 
Taken as a whole, the results indicate that nuclear effects in the charges are typically at the $\lsim$2\% level in light nuclei with atomic number $A\le 3$. The exception to this picture is in the scalar channel where $\sim10$\% effects are seen. 
For each type of interaction, 
nuclear modifications scale approximately with the magnitude of the corresponding charge.
While strange quark (equivalently, disconnected) contributions to the nuclear axial and tensor charges are negligible, strange quarks make significant contributions to the scalar charges, as seen for matrix elements of the same operators in the proton in previous studies~\cite{Green:2015wqa,Bhattacharya:2015wna,Alexandrou:2017qyt}.

The tensor charges encode the quark EDM contributions to the EDMs of light nuclei and thus set bounds on BSM sources of CP violation~\cite{Chupp:2017rkp}. Given that the CP violation in the weak interaction is insufficient to generate the observed matter-anti-matter asymmetry of the universe 
(assuming exact CPT invariance and baryon--anti-baryon symmetry of the initial conditions), many experiments have sought to measure permanent EDMs as evidence for such sources. 
Even with a successful measurement of a permanent EDM, fully disentangling the sources of CP violation requires multiple observables~\cite{Engel:2013lsa,Mereghetti:2015rra}, 
and experiments searching for EDMs of light nuclei are in the planning stages~\cite{Semertzidis:2003iq,Semertzidis:2011qv,Pretz:2013us}.
Nuclear effects in the tensor charge have not been previously observed; here they are resolved for the first time and found to be at the few percent level for $A\leq3$ at these quark masses. 
Similarly, modification of the axial charge in nuclei is found to be at the 1--2\% level for both the isoscalar and isovector combinations. 
The isovector $^3$He charge is consistent with values extracted from measurements of the $\beta$ decay of tritium \cite{Hardy:2014qxa} and is more precise than our previous work \cite{Savage:2016kon}.  
Nuclear effects in the axial charges can test predictions that nuclear modification of the spin-dependent 
structure function may be significantly different than the modification of the spin-independent structure function~\cite{Chen:2004zx,Cloet:2005rt,Smith:2005ra}. 
The small deviation resolved in this study implies that quarks in nuclei carry a  different fraction of the total spin than 
quarks in free nucleons. 

In contrast to the few-percent nuclear effects seen in the tensor and axial charges, the scalar charges of light nuclei are suppressed at the 10\% level relative to expectations for non-interacting nucleons.\footnote{The sign of these nuclear effects is consistent with the deeper binding of nuclei with increasing quark masses that is found from direct calculations of the binding energies of light nuclei~\cite{Beane:2012vq}.}  
In phenomenological models of nuclei such as the Walecka model~\cite{Walecka:1974qa,Serot:1984ey} and the 
quark-meson coupling model~\cite{Saito:1994ki}, a mean scalar field in which the nucleons move is an important contribution 
to the saturation of nuclear matter. The large modifications of the scalar charges found here 
suggest that models  based on similar mechanisms may approximately describe nuclei even at  unphysical values of the quark masses. 
A determination of the scalar polarizabilities through extensions of the calculations presented here (using analogues of the methods discussed in Refs.~\cite{Chang:2015qxa,Shanahan:2017bgi,Tiburzi:2017iux}) would be interesting in this context \cite{Guichon:1987jp,Stone:2016qmi}.

The scalar charges of nuclei are also important in the interpretation of experimental searches for  dark matter~\cite{Engel:1992bf,Prezeau:2003sv,Ellis2008,Ellis2009,Giedt2009,Hill2011,Ellis2012,Underwood2012,Korber:2017ery,Cirigliano:2012pq,Fitzpatrick:2012ix,Klos:2013rwa,Hoferichter:2015ipa,Bishara:2016hek,Hoferichter:2017olk}. 
These charges quantify the contribution of explicit chiral symmetry breaking to nuclear masses~\cite{Birse1994,Beane:2013kca}, 
and define nuclear {\it $\sigma$-terms}.
The $\sigma$-terms govern the interaction probabilities of many particle dark matter candidates with nuclei in direct detection experiments.
The pion and strange $\sigma$-terms for a nucleus $A$ are defined in analogy to the nucleon $\sigma$-terms as:
$\sigma_{\pi A} = m_l \langle A | \overline{u} u + \overline{d} d | A \rangle$ and 
$\sigma_{sA} = m_s \langle A | \overline{s} s | A \rangle$, respectively (where $m_l$ denotes the average light-quark mass). They can be determined from the scalar charges calculated here and are tabulated in Table~\ref{tab:sigmas}.
\begin{table}
	\begin{ruledtabular}
		\begin{tabular}{c|cccc}
			& $p$ & $d$ & $pp$ & $^3$\text{He} \\ \hline
			$\sigma_{\pi A}$ (MeV) & 327(5) & 648(10) & 649(10) & 942(16) \\
			$\sigma_{sA}$ (MeV) & 22.5(8) & 43.4(19) & 43.5(18) & 60.7(29) \\ \hline
			$\sigma_{B; \pi A}$ (MeV)& - & -7 (14) & -6(15) & -40(22) \\
			$\sigma_{B; sA}$ (MeV) & - & -1.5(2.4) & -1.4(2.4) & -6.7(3.7)
		\end{tabular}
	\end{ruledtabular}
\caption{\label{tab:sigmas} 
The $\sigma$-terms of the proton and nuclei at the  quark masses used in this work. 
$\sigma_{B; \pi A}$ and $\sigma_{B; sA}$ denote the differences between the sigma-terms of nuclei and $A$ times the sigma-term of the proton~\cite{Beane:2013kca}. Here $m_s=96$ MeV has been used, consistent with the physical strange quark mass. 
}
\end{table}
The results for the light-quark $\sigma$-terms are consistent with, and  more precise than, those deduced from numerical lattice QCD 
calculations for these nuclei using a discretized Feynman-Hellmann approach~\cite{Beane:2013kca}.

\vspace*{3mm}
{\it Summary:} 
The  results of the 
lattice QCD calculations presented here reveal percent-level nuclear effects in the axial and tensor charges of light nuclei, 
and ${\cal O}(10\%)$  nuclear effects in the scalar charges,  at unphysical values of the quark masses.  
This  is consistent with nucleons being the  relevant effective degrees of freedom within these light nuclei, dominating 
nuclear responses to external probes.
Future calculations using additional lattice spacings and volumes, and with the physical values of the quark masses, will determine the scalar, axial and tensor matrix elements of light nuclei with fully controlled uncertainties.
These  can then 
be used to constrain EFT analyses of electroweak interactions with light nuclei and of searches for BSM physics. 
Specifically, the tensor charges of nuclei are needed for the interpretation~\cite{Engel:2013lsa,Mereghetti:2015rra,Wirzba:2016saz} of potential future experimental searches for nuclear EDMs~\cite{Semertzidis:2003iq,Semertzidis:2011qv,Pretz:2013us}, and
the axial charges control Standard Model predictions for GT transitions and double-$\beta$ decay rates.
The scalar charges enter the interpretation of dark matter direct detection experiments and searches for new physics in precision spectroscopy.
If the significant nuclear modifications to scalar matrix elements observed in this work  persist in larger nuclei in nature, such effects  will  be important for the interpretation of intensity-frontier searches for new physics that employ nuclear targets.

\vspace*{2mm}

{\it Acknowledgments:}
We thank Silas Beane, Martin Hoferichter, 
Bob McKeown, Assumpta Parre\~no, Yotam Soreq, Brian Tiburzi and Nodoka Yamanaka for helpful discussions and comments. 
We would like to thank Jordy de Vries for emphasizing to us that it may be possible to understand the trends seen in these observables
from the nature of the nuclear forces  at this pion mass.
This research was supported in part by the National Science Foundation under grant number NSF PHY11-25915 and ZD, WD, MJS, PES and MLW acknowledge the Kavli Institute for Theoretical Physics for hospitality during the development of this work.
Calculations were performed using computational resources provided
by  NERSC (supported by U.S. Department of Energy grant number DE-AC02-05CH11231), and by the USQCD collaboration.  This research used resources of the Oak Ridge Leadership Computing Facility at the Oak Ridge National Laboratory, which is supported by the Office of Science of the U.S. Department of Energy under Contract number DE-AC05-00OR22725. We acknowledge use of the College of William and Mary computing facilities supported by NSF (MRI grant PHY-1626177), the Commonwealth of Virginia Equipment Trust Fund and the Office of Naval Research.
The PRACE Research Infrastructure resources  at the Tr\`es Grand Centre de Calcul and Barcelona Supercomputing Center were also used.
Parts of the calculations used the Chroma software suite~\cite{Edwards:2004sx}.  
ZD was partly supported by the Maryland Center for Fundamental Physics.
WD was partly supported by  U.S. Department of Energy Early Career Research Award DE-SC0010495 and grant number DE-SC0011090.
 AG is supported under the auspices of the U.S. Department of Energy by LLNL under Contract No. DE-AC52-07NA27344.
KO was partially supported by the U.S. Department of Energy through grant
number DE- FG02-04ER41302 and  by STFC consolidated grant ST/P000681/1. KO, PES and FW were partially supported through contract number DE-AC05-06OR23177 under which JSA operates the Thomas Jefferson National Accelerator Facility.  
MJS was supported by DOE grant number~DE-FG02-00ER41132, and in part by the USQCD SciDAC project, the U.S. Department of Energy through grant number DE-SC00-10337.	
MLW was supported by a MIT Pappalardo Fellowship and in part by  DOE grant number~DE-SC0011090.
FW was partially supported through the USQCD Scientific Discovery through Advanced Computing (SciDAC) project funded by U.S. Department of Energy, Office of Science, Offices of Advanced Scientific Computing Research, Nuclear Physics and High Energy Physics.

\bibliography{charges}

\appendix

\section*{Supplementary Material}

\begin{figure}[H]
	\centering 
	\includegraphics[width=\columnwidth]{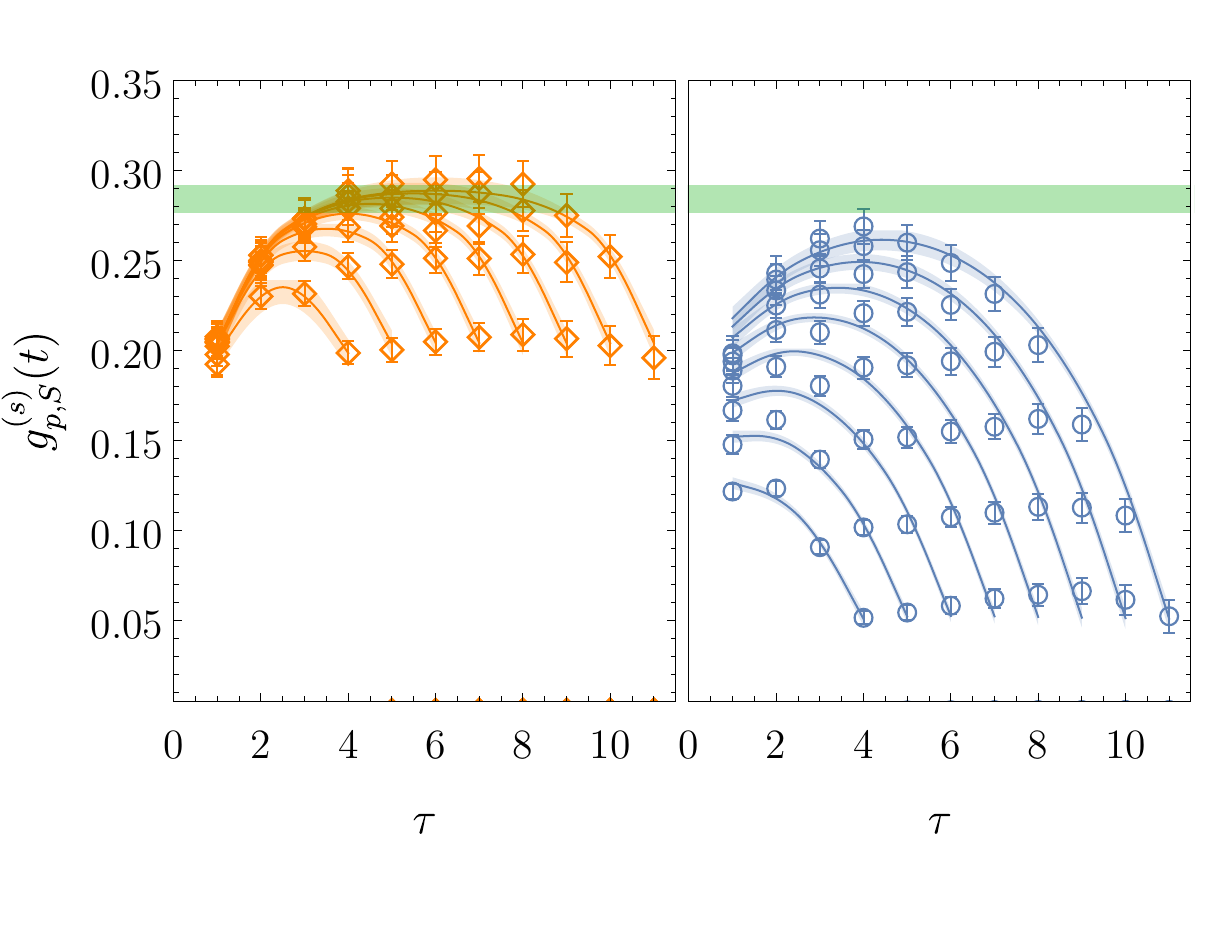}
	\includegraphics[width=\columnwidth]{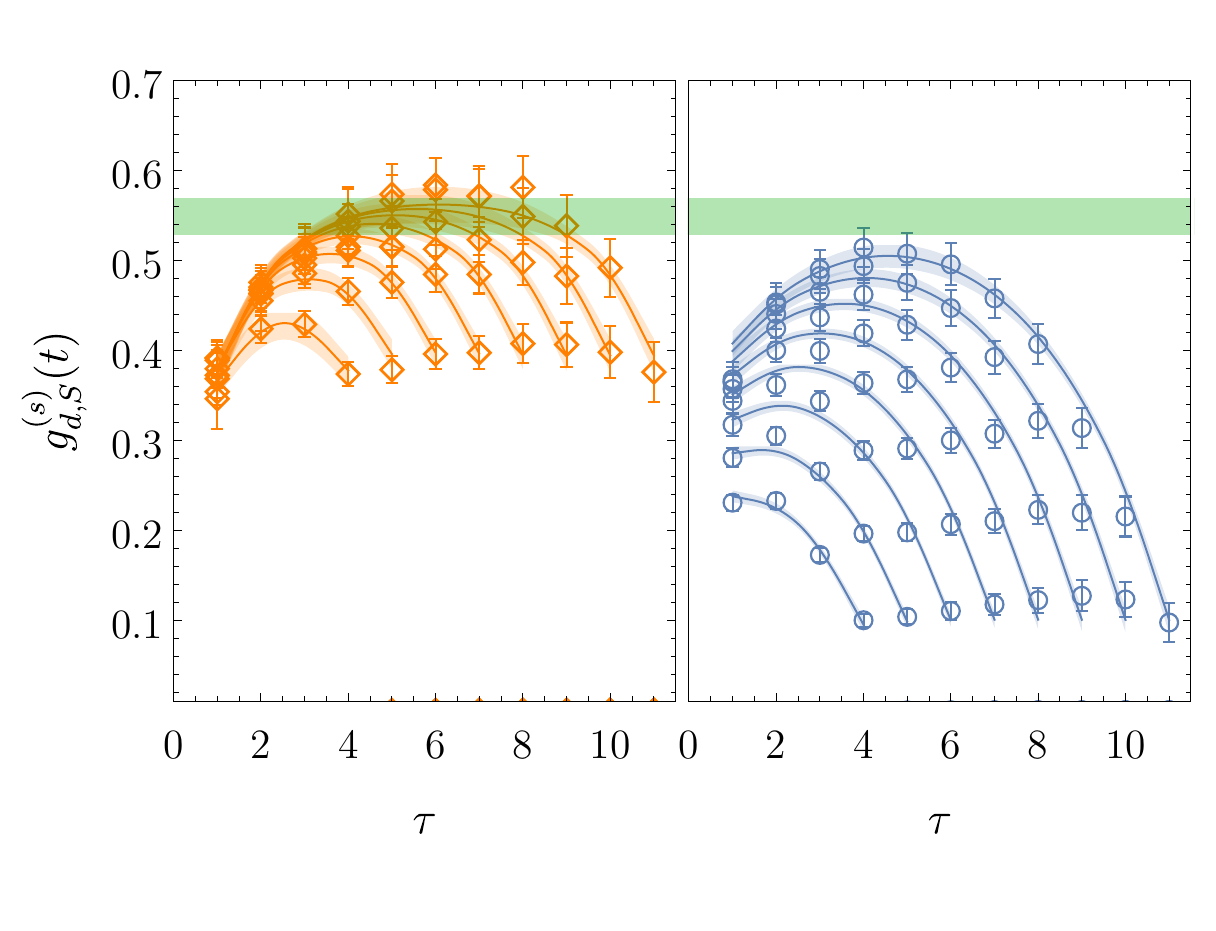}
	\includegraphics[width=\columnwidth]{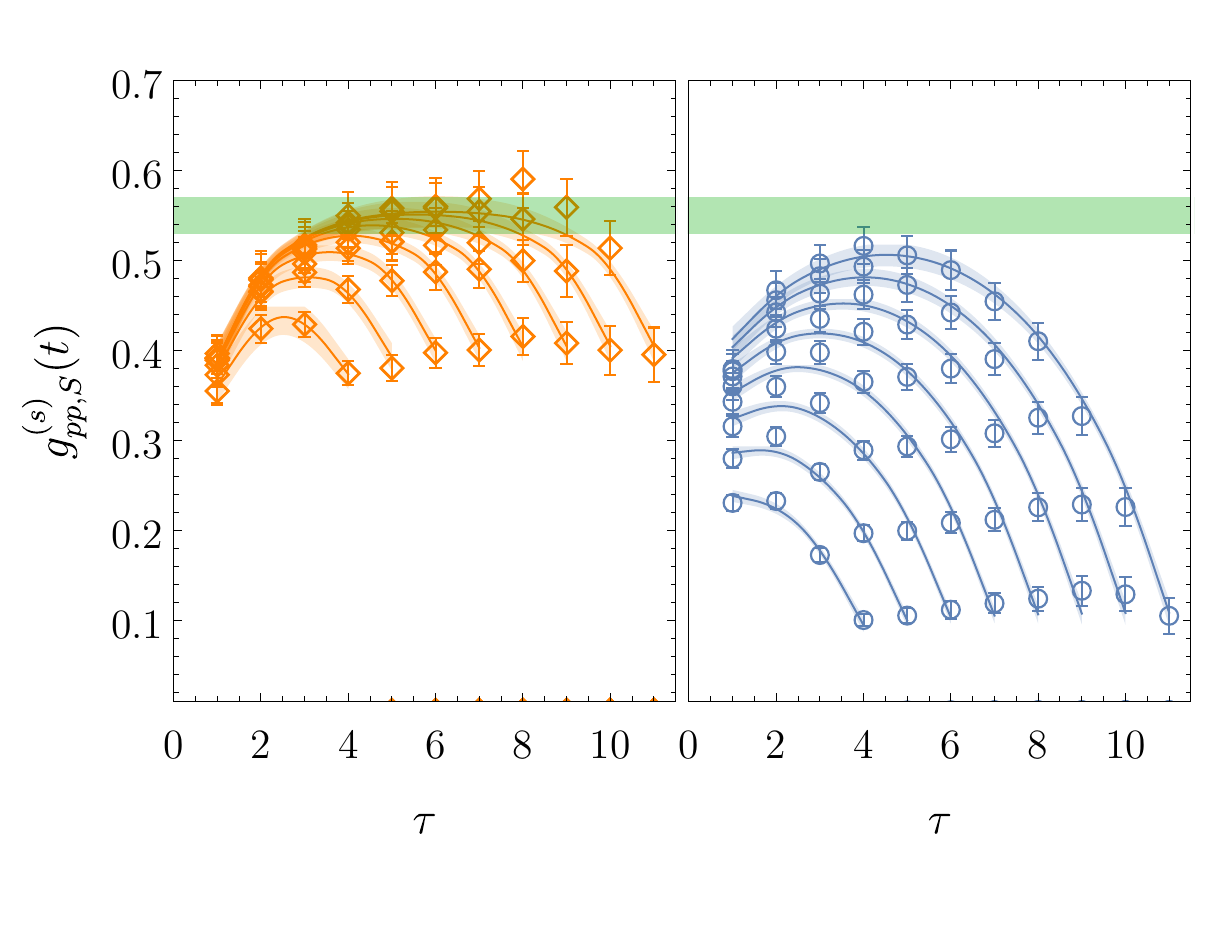}
	\caption{\label{fig:discoextra} The strange quark (disconnected) contribution to the scalar charges in the proton (top), deuteron (middle) and diproton (bottom). The left (right) panels show the SS (SP) correlation functions as a function of the current insertion time $\tau$ for a range of sink  times. The green band corresponds to the extracted charge and the blue (orange) curves and bands denote a coupled two-state fit to all data shown.}
\end{figure}

\begin{figure}[H]
	\centering
	\includegraphics[width=\columnwidth]{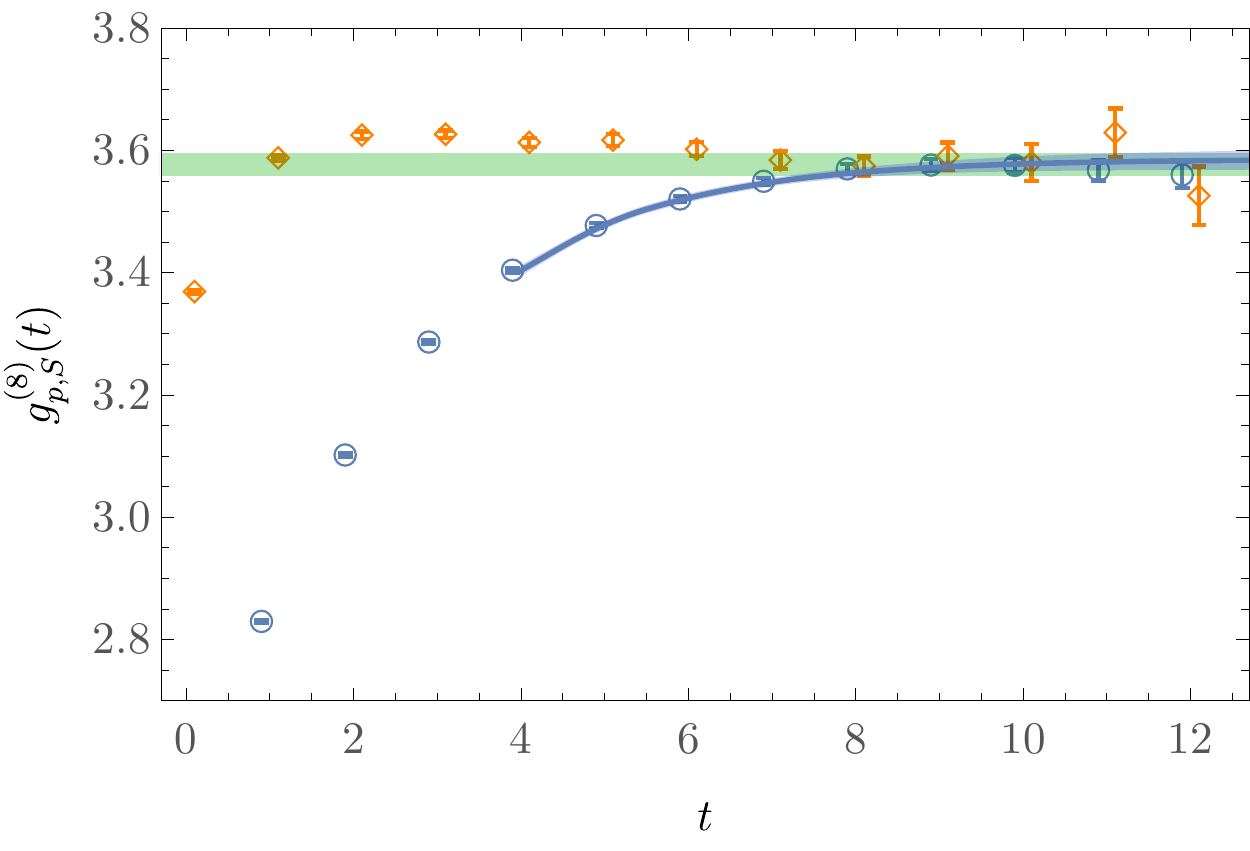}
	\includegraphics[width=\columnwidth]{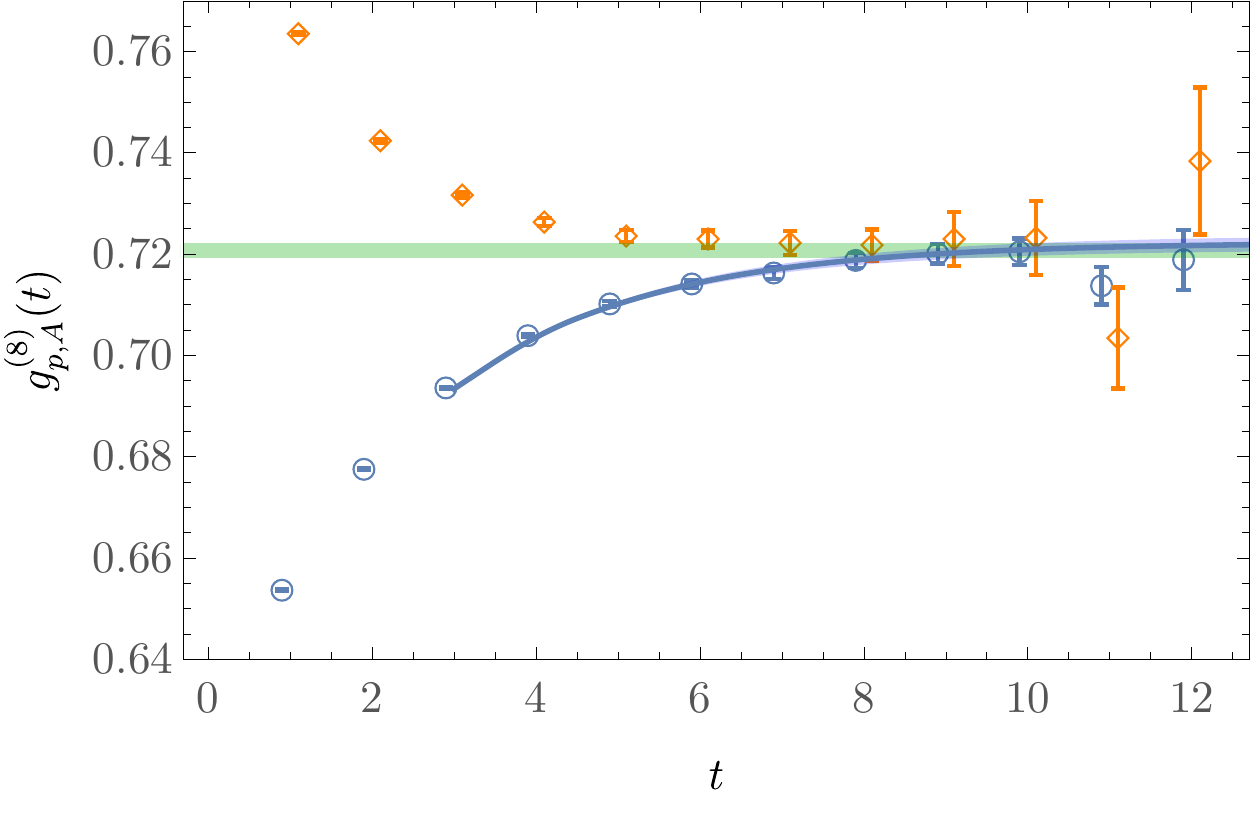}
	\includegraphics[width=\columnwidth]{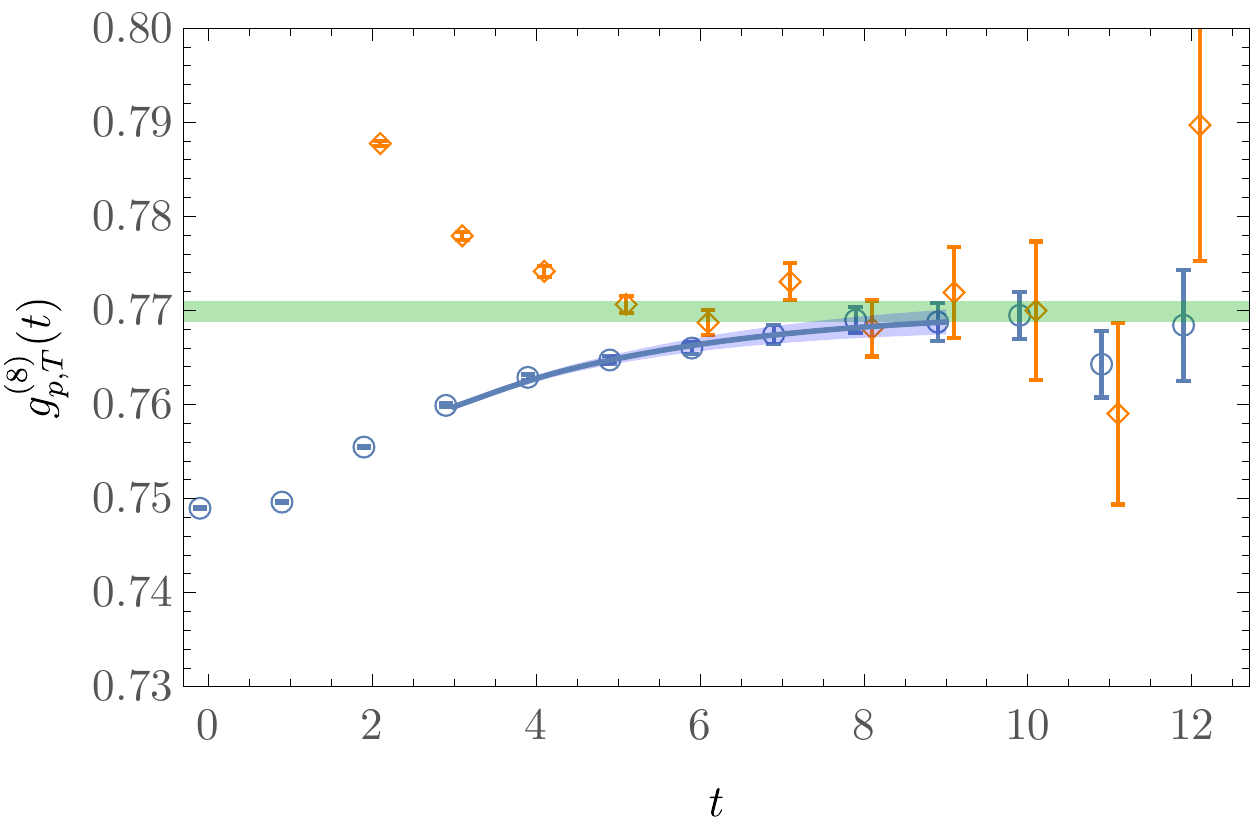}
	\caption{\label{fig:nucleonisoscalarconnected} Bare effective charges for the connected proton isoscalar scalar (top), axial (middle) and tensor (bottom) connected contractions. The blue circles and orange diamonds denote SP and SS results, respectively.}
\end{figure}

\begin{figure}[H]
	\centering
	\includegraphics[width=\columnwidth]{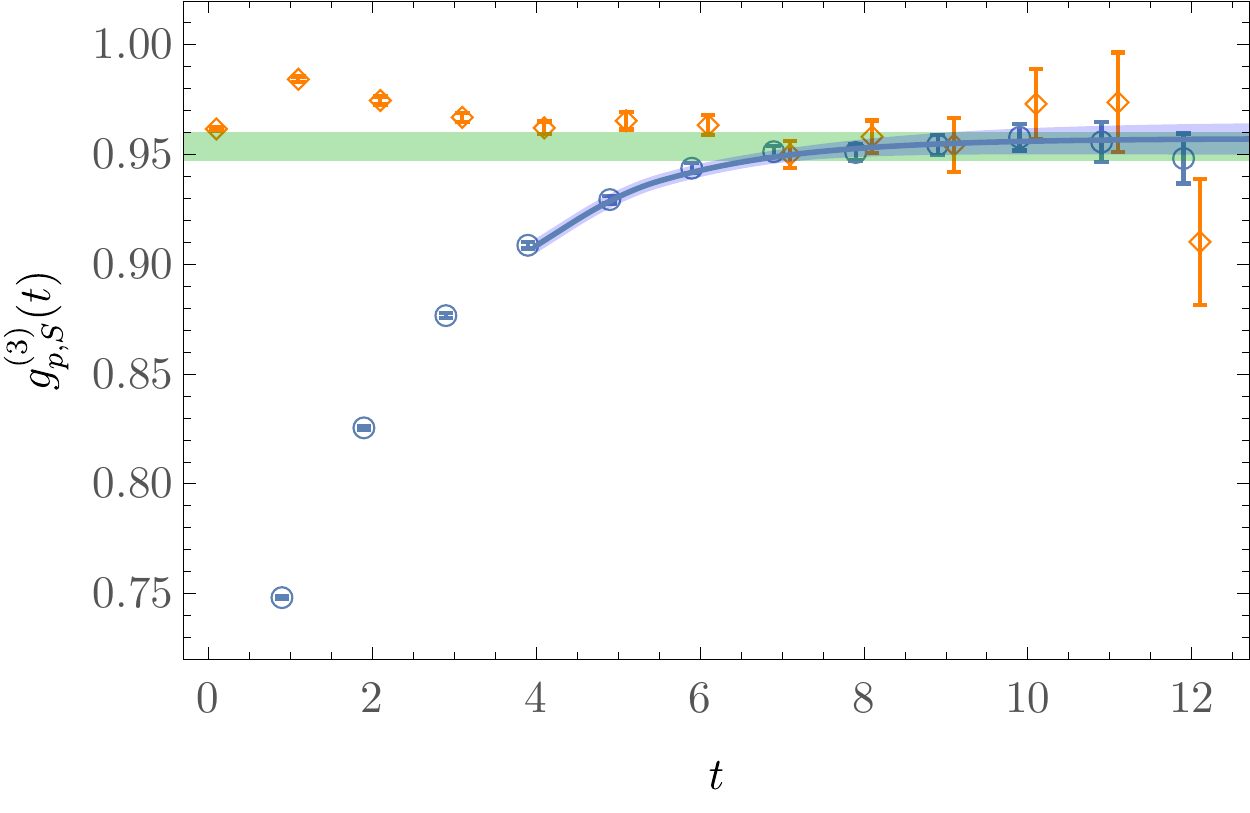}
	\includegraphics[width=\columnwidth]{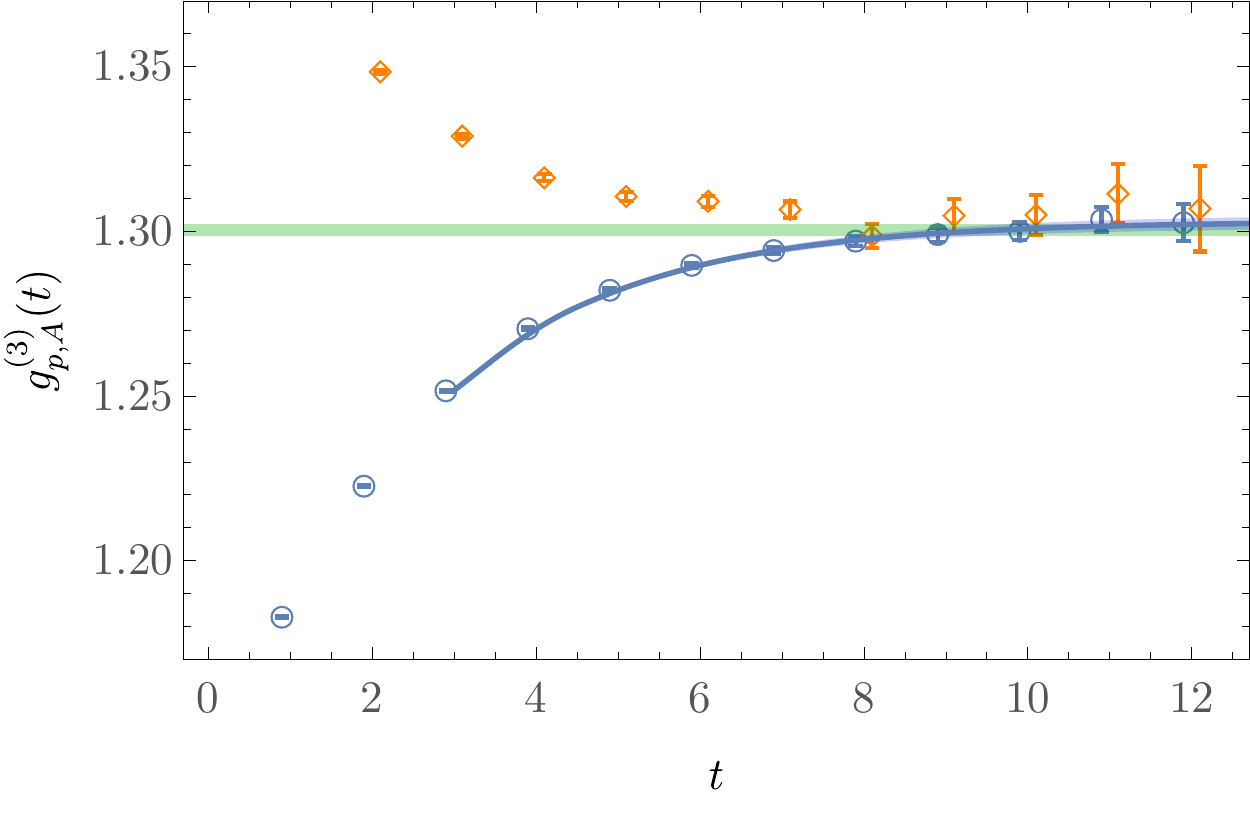}
	\includegraphics[width=\columnwidth]{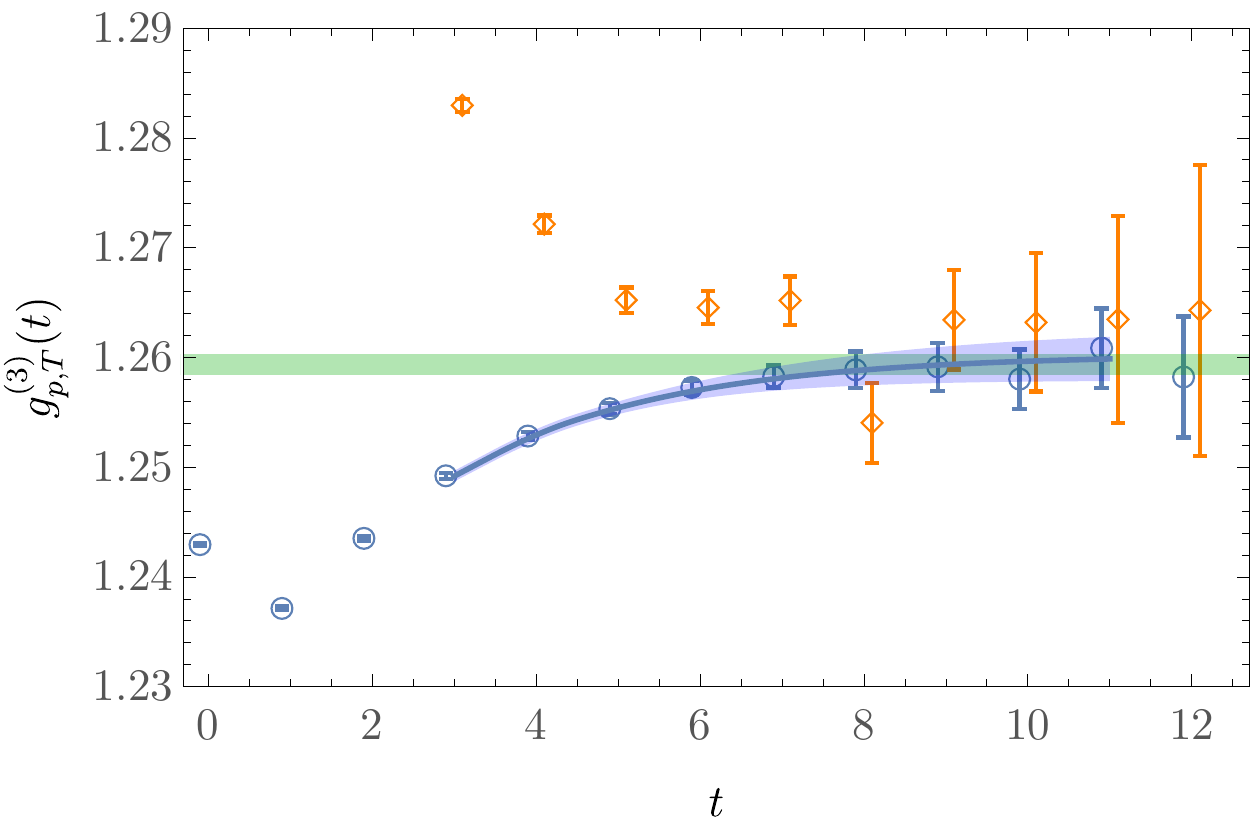}
	\caption{\label{fig:nucleonisovectorconnected} Bare effective charges for the connected proton isovector scalar (top), axial (middle) and tensor (bottom) connected contractions. The blue circles and orange diamonds denote SP and SS results, respectively.}
\end{figure}

\begin{figure}[H]
	\centering
	\includegraphics[width=\columnwidth]{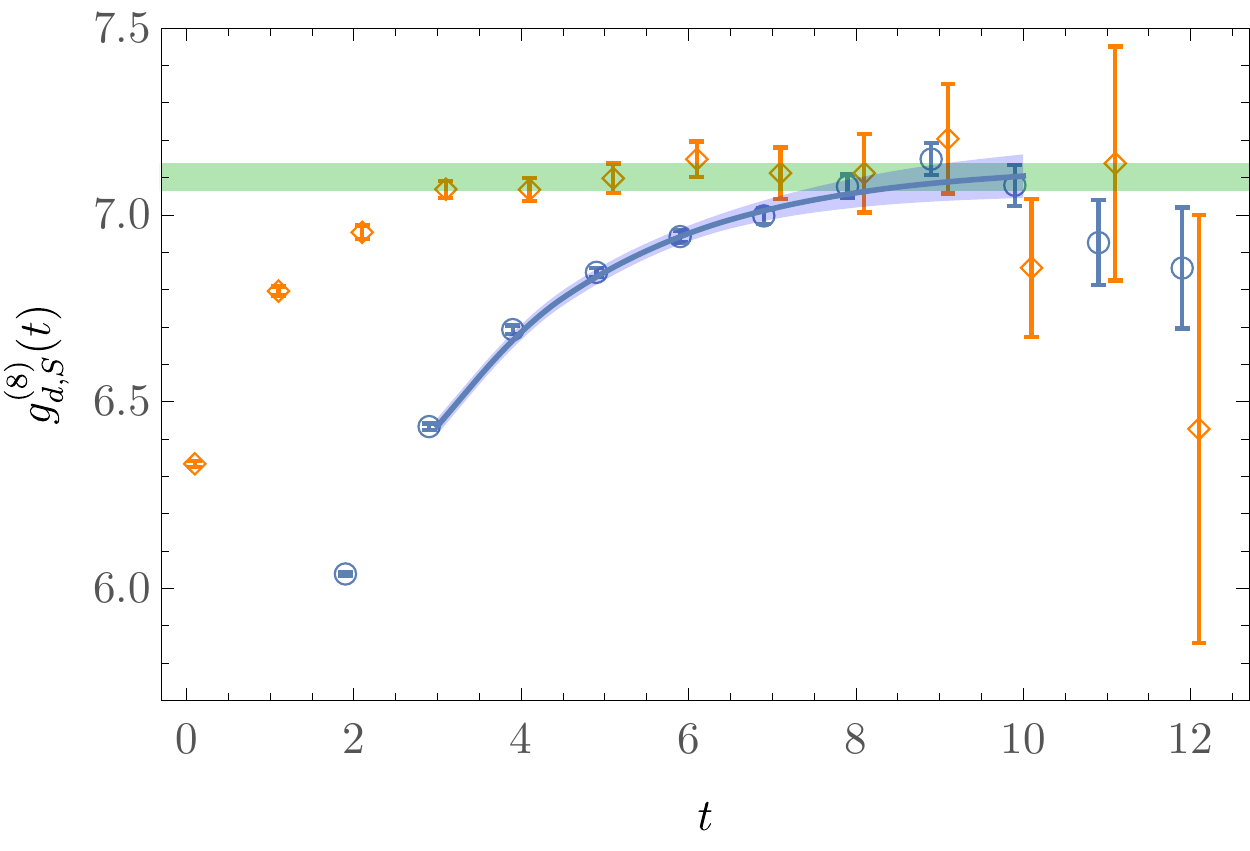}
	\includegraphics[width=\columnwidth]{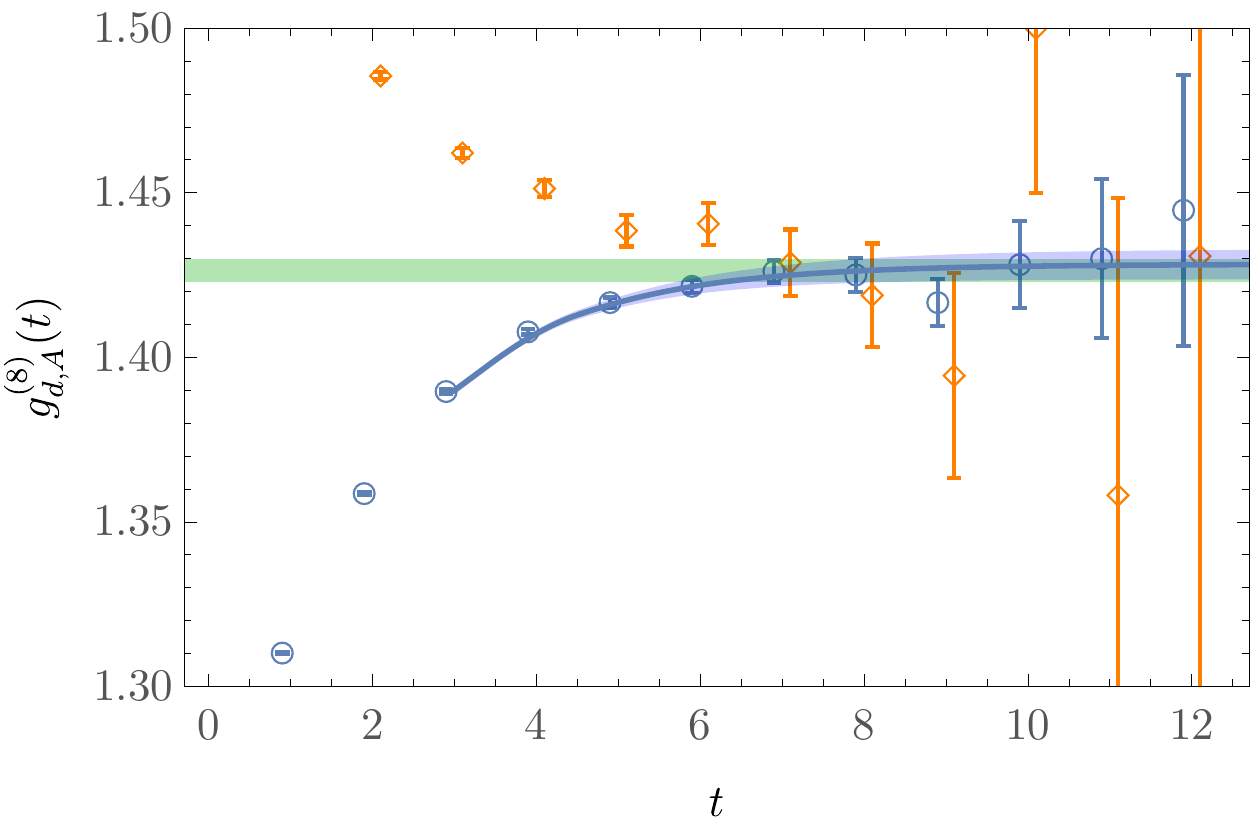}
	\includegraphics[width=\columnwidth]{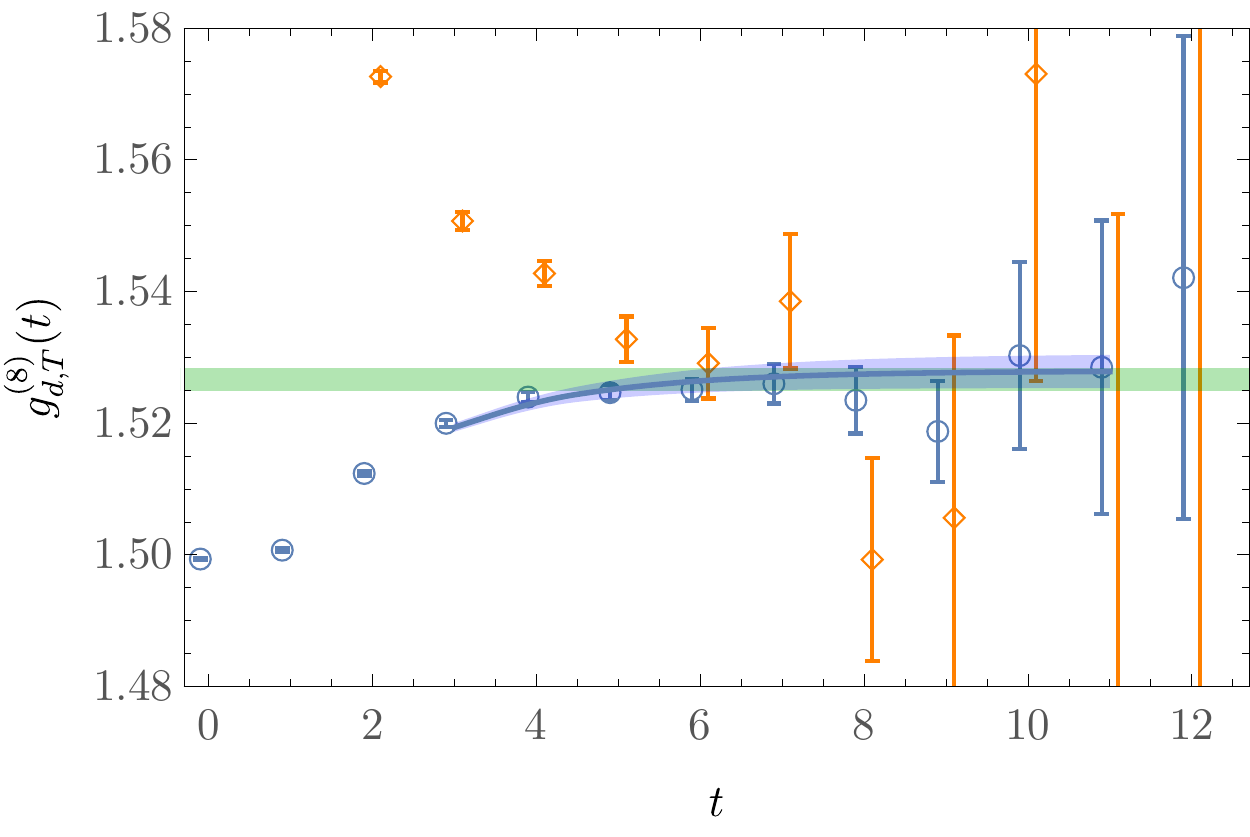}
	\caption{\label{fig:deuteronconnected} Bare effective charges for the connected deuteron isoscalar scalar (top), axial (middle) and tensor (bottom) connected contractions. The blue circles and orange diamonds denote SP and SS results, respectively.}
\end{figure}

\begin{figure}[H]
	\centering
	\includegraphics[width=\columnwidth]{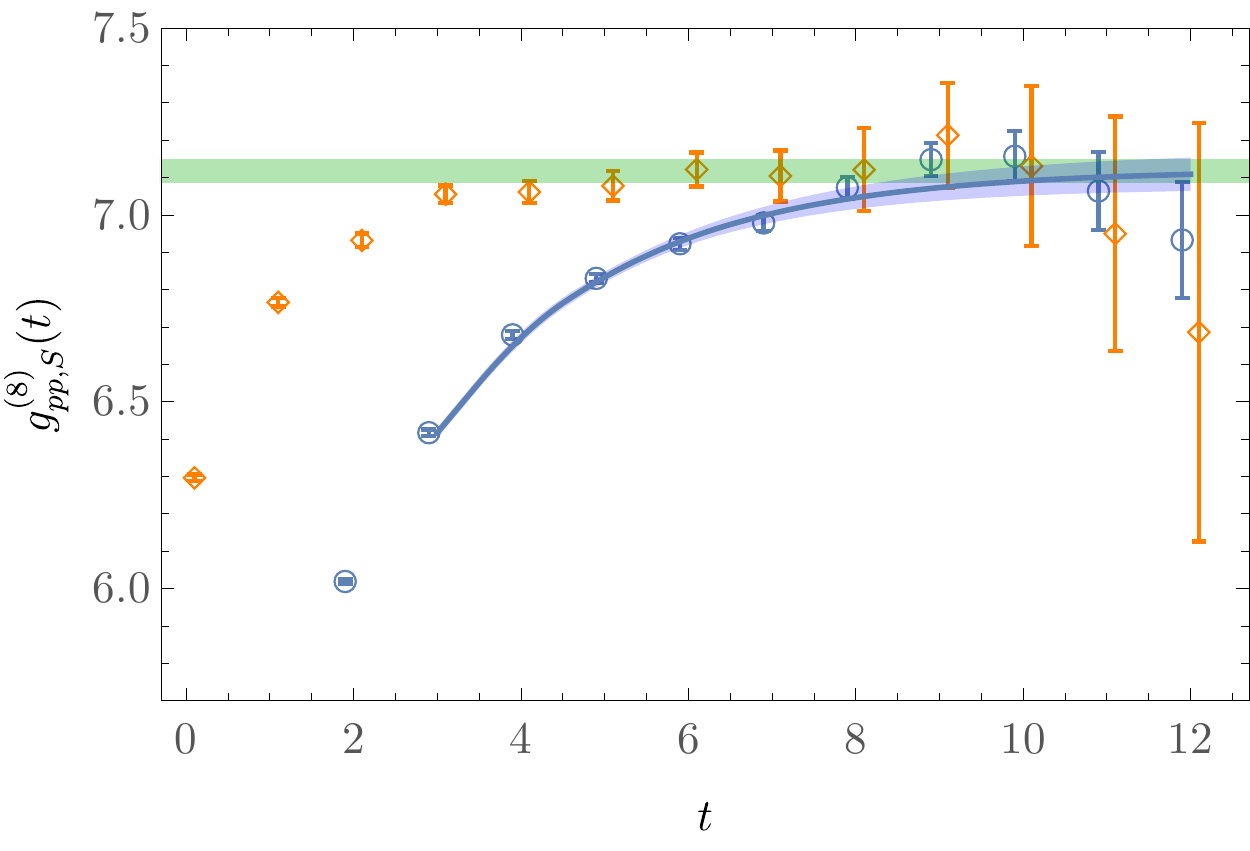}
	\includegraphics[width=\columnwidth]{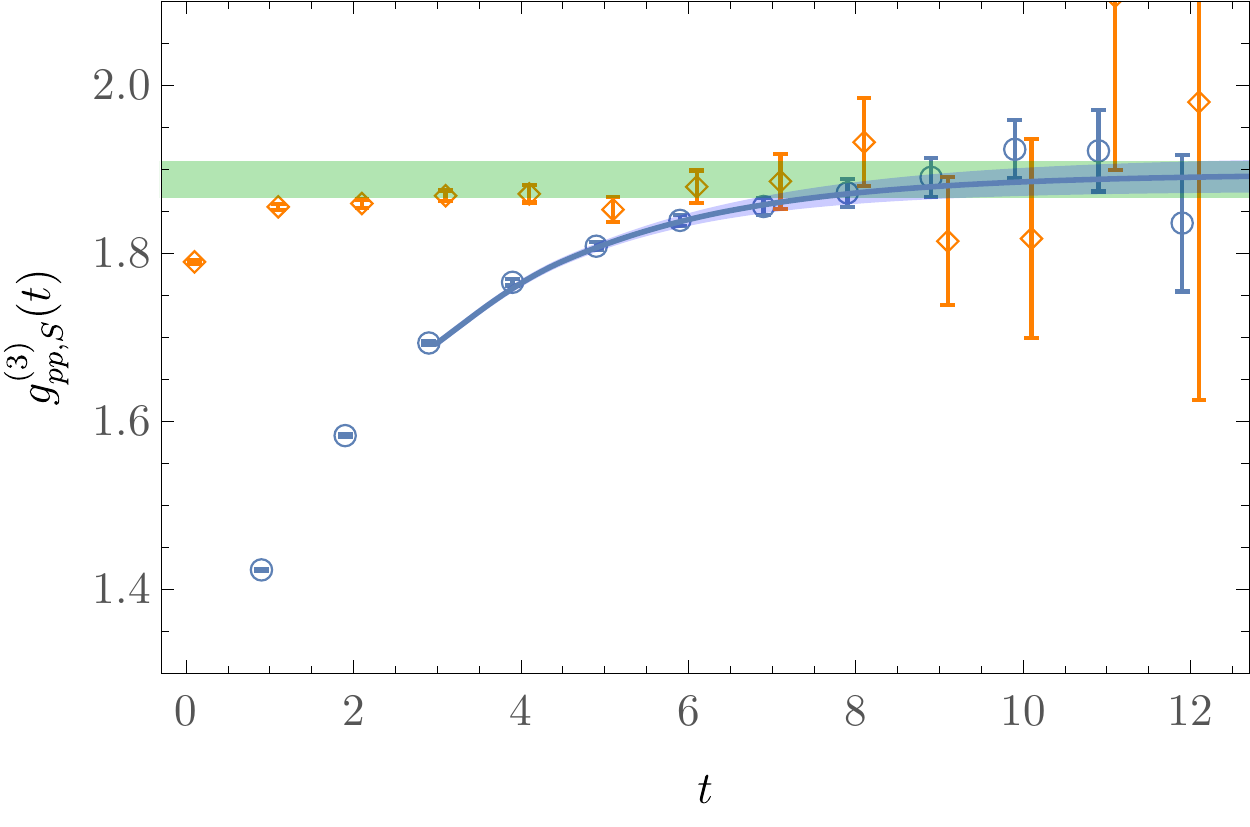}
	\caption{\label{fig:dinucleonconnected} Bare effective charges for the connected diproton isoscalar (top) and isovector (bottom) scalar connected contractions. The blue circles and orange diamonds denote SP and SS results, respectively.}
\end{figure}

\begin{figure}[H]
	\centering
	\includegraphics[width=\columnwidth]{TritScalarIsoscalar.pdf}
	\includegraphics[width=\columnwidth]{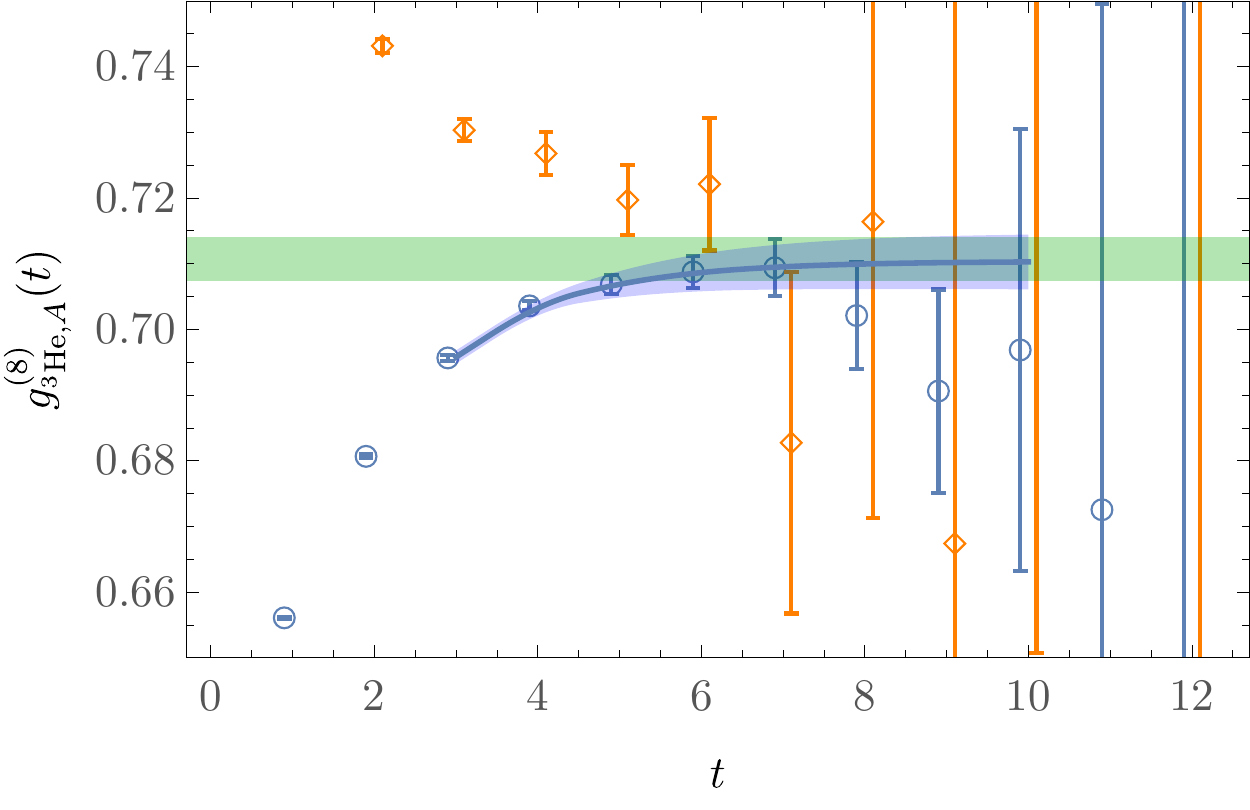}
	\includegraphics[width=\columnwidth]{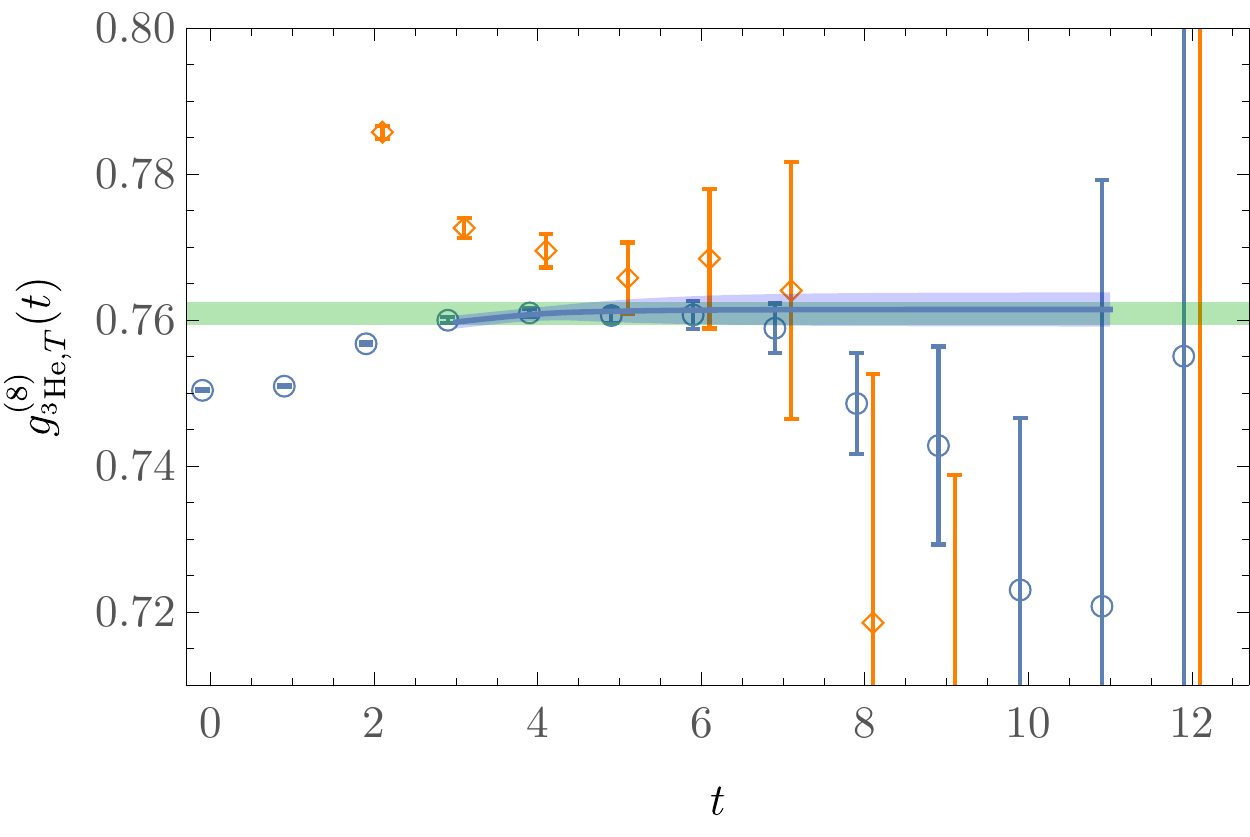}
	\caption{\label{fig:tritonisoscalarconnected} Bare effective charges for the connected $^3$He isoscalar scalar (top), axial (middle) and tensor (bottom) connected contractions. The blue circles and orange diamonds denote SP and SS results, respectively.}
\end{figure}

\begin{figure}[H]
	\centering
	\includegraphics[width=\columnwidth]{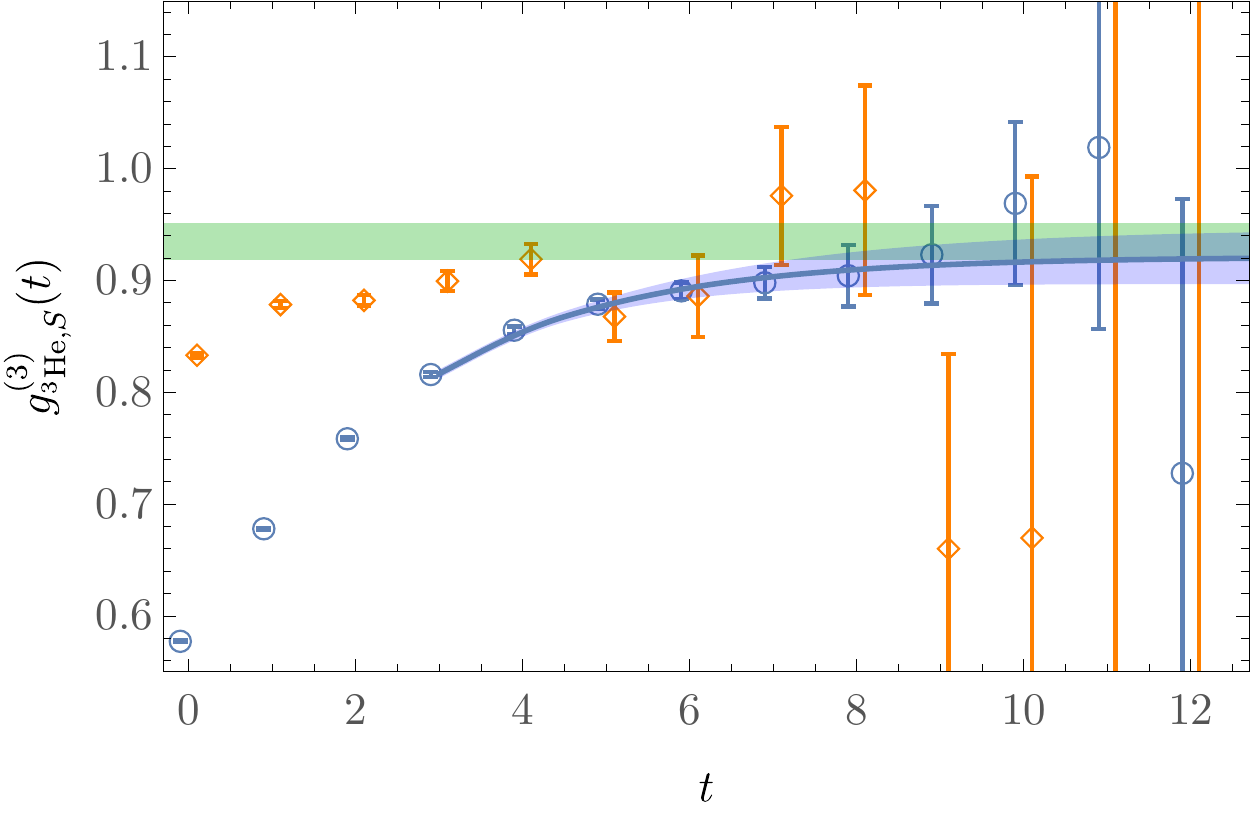}
	\includegraphics[width=\columnwidth]{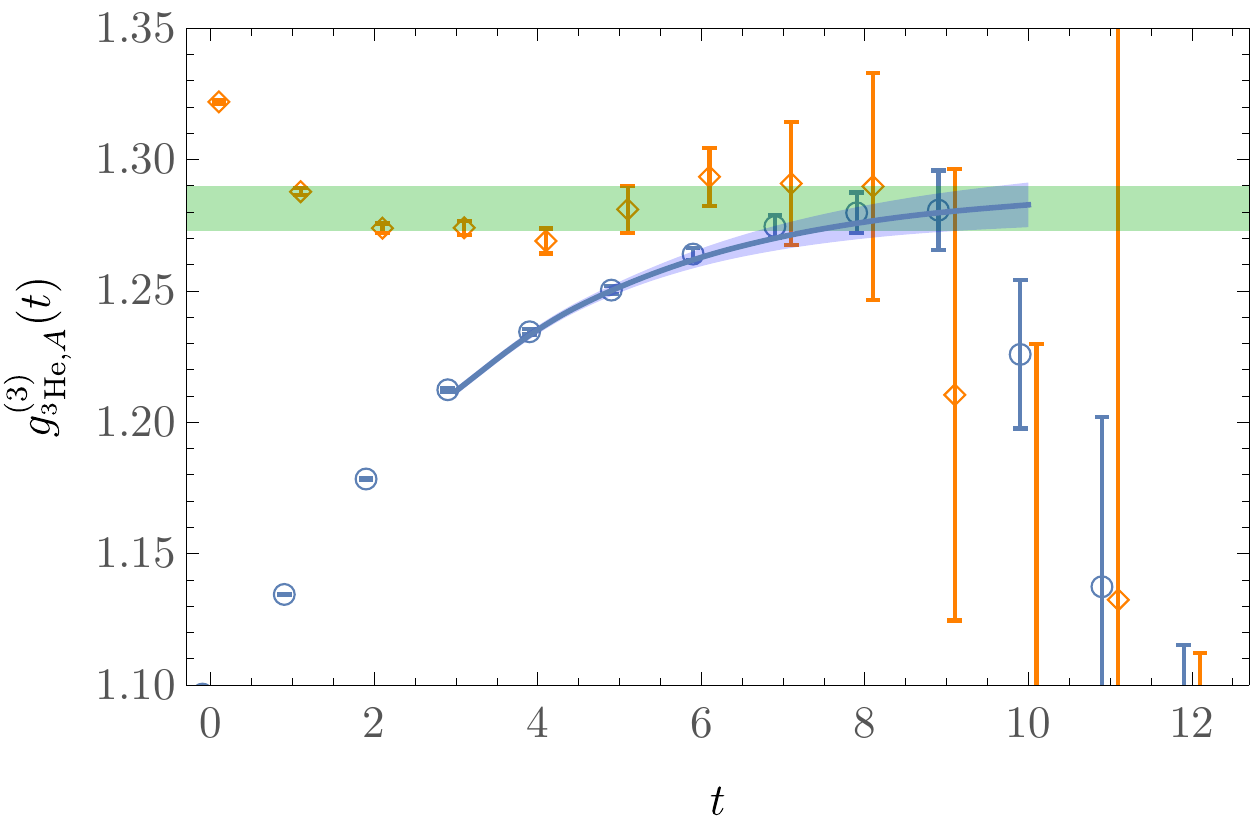}
	\includegraphics[width=\columnwidth]{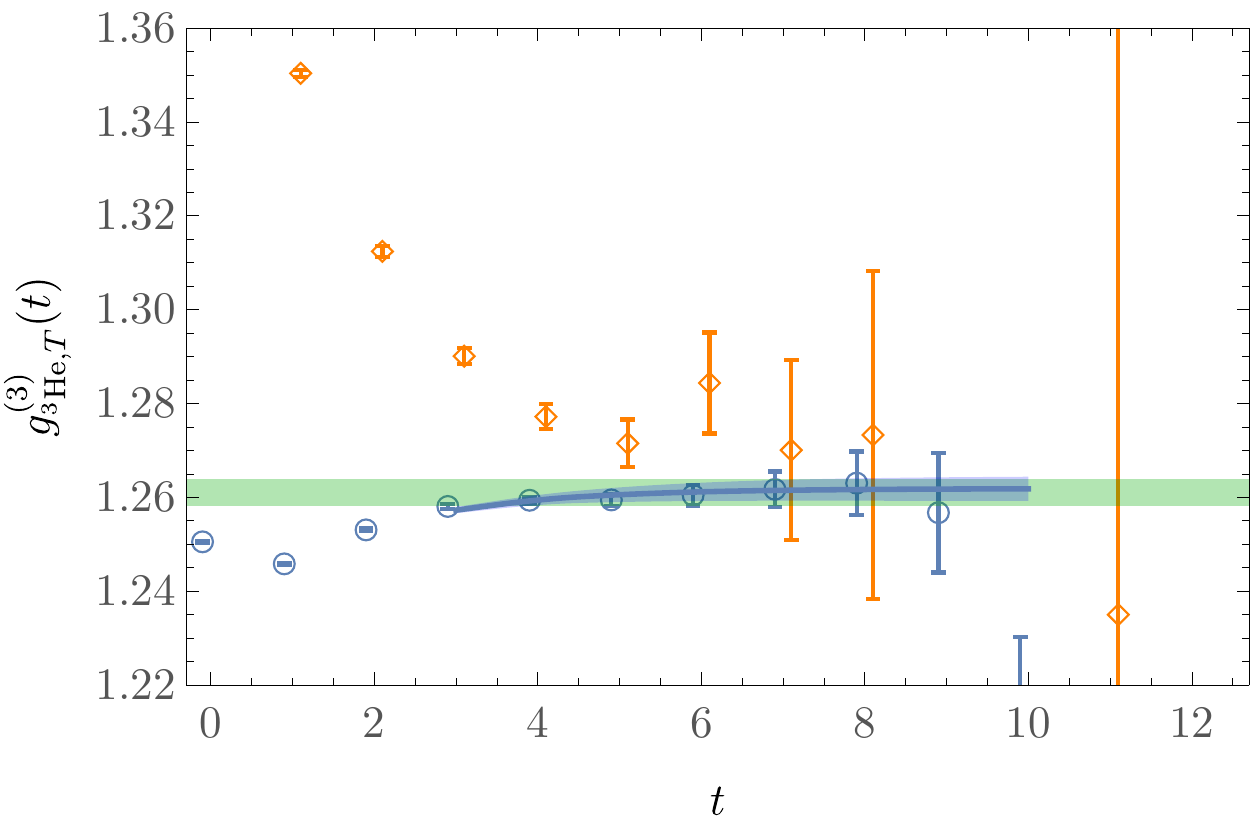}
	\caption{\label{fig:tritonisovectorconnected} Bare effective charges for the connected $^3$He isovector scalar (top), axial (middle) and tensor (bottom) connected contractions. The blue circles and orange diamonds denote SP and SS results, respectively.}
\end{figure}

The statistical samplings used in these calculations are summarized in Table \ref{tab:stats}. 
The effective charges for connected contributions, and the three-point to two-point correlation function ratios for disconnected contributions, are illustrated in Figs.~\ref{fig:discoextra}--\ref{fig:tritonisovectorconnected}.
In combination with Figs.~1 and 2 of the main text, these show all
Dirac and flavor structures for all states investigated in this work.
{\color{black} A sample two-state fit to the SP correlators is shown on each figure in the connected cases, with a fit form 
	\begin{equation}
	g(t)=A + B \ e^{-\Delta t}\,,
	\end{equation}
	where $A$, $B$ and $\Delta$ are fit parameters. Similarly, a correlated simultaneous two-state fit to the SS and SP ratios is shown in the disconnected cases, with dependence on the two temporal coordinates
	of the form
	\begin{equation}
	g(t,\tau)=A + B e^{-\Delta t} + C e^{-\Delta(t-\tau)} + D e^{-\Delta \tau}\,,
	\end{equation}
	where $A$, $B$, $C$, $D$ and $\Delta$ are fit parameters ($B$, $C$ and $D$ are allowed to vary between the SS and SP data, but $A$ and $\Delta$ are common). This form arises from the consideration of the ratios of two and three point correlation functions. }
The green bands show the final results, and include both statistical uncertainties and systematic uncertainties arising from variations of the fit window (in both $t$ and $\tau$ in the disconnected case) and fit form (1 or 2-state fits). Several independent analyses were performed in each case, including additional approaches such as single-state and summation-method analyses for the disconnected matrix elements, and consistent results were found for both the central values and uncertainties of the matrix elements. Despite considerably smaller statistical sampling, the connected contributions are determined to  higher precision 
than the disconnected contributions due to the larger magnitudes of the connected contributions.

\begin{table}[!h]
	\begin{ruledtabular}
		\begin{tabular}{c|ccccc}
			Channel & $N_{\rm cfg}$ & $N_{\rm src}$ & $N_{\rm HP}$ & $N_{\rm def}$\\ 
			\hline
			Scalar Connected & 1018 & 16 & -- & -- \\
			Axial Connected & 662 & 16 & -- & -- \\
			Tensor Connected & 662 & 16 & -- & -- \\
			Disconnected & 508 & 416 & 256 &  500 & \\
		\end{tabular}
	\end{ruledtabular} 
	\caption{Details of the measurements performed for the  charges. 
		For the connected calculations, $N_{\rm src}$ sources were used on $N_{\rm cfg}$ configurations 
		separated by 10 hybrid Monte Carlo updates. For the disconnected calculations, $N_{\rm HP}$ hierarchical probing vectors were used with full spin-color dilution and  $N_{\rm def}$ singular triplets \protect\cite{Gambhir:2016uwp} were used for deflation in the estimation of the disconnected loop ${\rm tr}[(D\!\!\!\! / +m)^{-1}\Gamma]$ for each Dirac structure $\Gamma=1,\ \gamma_3\gamma_5, \ \gamma_1\gamma_2$. These loops were then correlated with nuclear two-point functions calculated from $N_{\rm src}$ source locations at temporal separations $\tau=1,\ldots,12$ from the loop insertion. 
		\label{tab:stats}}
\end{table}

For completeness, the bare nuclear charges are presented in Table \ref{tab:bare}, and the ratios shown in Fig.~3 of the main text are listed in Table \ref{tab:rat}. The bare charges are in many cases more precise than the renormalized charges shown in the main text as the renormalization factors are less precise than the matrix-element calculations. The expectations for collections of non-interacting nucleons are also shown.

\begin{table}[H]
	\begin{ruledtabular}
		\begin{tabular}{c|cccc}
			& $p$ & $d$ & $pp$ & ${}^3$He\\ 
			\hline
			$g_S^{(0)}$	&4.43(3) & 8.75(7) & 8.77(7) & 12.70(13)  \\ 
			$g_S^{(3)}$	&0.953(7) & - & 1.89(2) & 0.93(2)\\ 
			$g_S^{(8)}$	& 3.57(2) & 7.10(4) & 7.12(3) & 10.39(6)  \\ 
			$g_S^{(s)}$	& 0.284(8) & 0.55(2) & 0.55(2) & 0.77(3)  \\ 	\hline
			$g_A^{(0)}$	& 0.721(3) & 1.431(5) & - & 0.720(7)  \\ 
			$g_A^{(3)}$	& 1.300(2) & - & - & 1.281(9) \\ 
			$g_A^{(8)}$	& 0.7201(14) & 1.426(4) & - & 0.711(3)  \\ 
			$g_A^{(s)}$	& 0.0002(6) & 0.0015(10) & - & 0.003(2)  \\ \hline
			$g_T^{(0)}$	& 0.7700(12) & 1.527(2) & - & 0.762(2)  \\ 
			$g_T^{(3)}$	& 1.259(1) & - & - & 1.261(3)  \\ 
			$g_T^{(8)}$	& 0.770(1) & 1.527(2) & - & 0.761(2)  \\ 
			$g_T^{(s)}$	& 0.00002(10) & 0.0003(2) & - & 0.005(4)  \\ 
		\end{tabular}
	\end{ruledtabular} 
	\caption{The bare  scalar, axial and tensor charges of the proton and light nuclei. }
	\label{tab:bare}
\end{table}

\begin{table}[H]
	\begin{ruledtabular}
		\begin{tabular}{c|ccc|c}
			& $d$ & $pp$ & ${}^3$He & NSN\\ 
			\hline
			$R_S^{(0)}$	& 1.97(2) &1.98(2)&2.87(4)& $B$\\ 
			$R_S^{(3)}$	& - &1.98(2)&0.96(2)& $2T_3$\\ 
			$R_S^{(8)}$	& 1.98(1) &1.99(2)& 2.90(2)& $2B$\\ 
			$R_S^{(s)}$	& 1.93(9) &1.94(9)& 2.70(14)& $B$\\ 
			\hline
			$R_A^{(0)}$	& 1.98(1) & - & 0.999(6) & $2S_3$ \\ 
			$R_A^{(3)}$	& - & - & 0.987(4) & $4T_3S_3$\\ 
			$R_A^{(8)}$	& 1.983(4) &- & 0.990(3)  & $2S_3$\\ 
			$R_A^{(s)}$	& * &* &*  & $BS_3$\\ 
			\hline
			$R_T^{(0)}$	& 1.984(4) & - & 0.990(2) & $2S_3$\\ 
			$R_T^{(3)}$	& - & - & 1.002(2)&  $4T_3S_3$ \\ 
			$R_T^{(8)}$	& 1.986(5) & - & 0.991(3)  &  $2S_3$\\ 
			$R_T^{(s)}$	&*  &* & * & $BS_3$\\ 
		\end{tabular}
	\end{ruledtabular} 
	\caption{The ratios of the scalar, axial and tensor charges of light nuclei to those of the proton. 
		The column labeled NSN gives the value of each ratio using the naive single-nucleon approximation,  
		$S_3$ is the third component of spin, $T_3$ is the third component of isospin and $B$ is baryon number. 
		The asterisks (*) denote ratios in which both the numerator and denominator are consistent with zero and meaningful ratios could not be obtained.}
	\label{tab:rat}
\end{table}

\end{document}